\documentclass[authoryear,preprint,3p,onecolumn]{elsarticle}

\usepackage{booktabs}
\usepackage{amssymb,amsmath, amsfonts}
\usepackage{underscore}
\usepackage{multirow}
\usepackage{multicol}
\usepackage{float}
\usepackage{stfloats}
\usepackage{flushend}
\usepackage{wrapfig}
\usepackage{graphicx}
\usepackage{mwe}
\usepackage{subcaption}
\usepackage{caption}
\usepackage{threeparttable}
\usepackage[table]{xcolor}
\usepackage{array}
\usepackage{lineno}\modulolinenumbers[5]
\usepackage{blindtext}
\usepackage{changepage}
\usepackage[export]{adjustbox}
\usepackage{natbib}
\setcitestyle{authoryear,round}
\usepackage{hyperref}
\hypersetup{
	colorlinks=true,
	linkcolor=blue,
	citecolor=blue,
	urlcolor=blue,
	linktocpage=false,
	pdfborder={0 0 0},
}
\usepackage[nameinlink,capitalise]{cleveref}
\usepackage{geometry}
\usepackage{titlesec}
\titlespacing{\subsection}{0pt}{\baselineskip}{\baselineskip}
\titlespacing{\subsubsection}{0pt}{\baselineskip}{\baselineskip}

\renewcommand\toprule{\specialrule{1pt}{1pt}{1pt}}
\renewcommand\midrule{\specialrule{0.4pt}{0.4pt}{3pt}}

\journal{Information Processing and Management}

\begin{document}

\begin{frontmatter}

\title{CAWAL: A Novel Unified Analytics Framework for Enterprise Web Applications and Multi-Server Environments\tnoteref{tref}}
\tnotetext[tref]{This work was supported by Sakarya University Scientific Research Foundation (Project number: 2010-50-02-024).}

\author[a,b]{Özkan Canay\corref{cor1}}

\affiliation[a]{organization={Sakarya University},
	addressline={Institute of Natural Sciences, Dept. of Computer and IT Engineering},
	city={Serdivan},
	postcode={54050},
	state={Sakarya},
	country={Türkiye}}
\affiliation[b]{organization={Sakarya University of Applied Sciences},
	addressline={Vocational School of Sakarya, Dept. of Computer Tech.},
	city={Adapazari},
	postcode={54290},
	state={Sakarya},
	country={Türkiye}}
\cortext[cor1]{Corresponding author}
\ead{canay@subu.edu.tr}

\author[c]{Ümit Kocabıçak}
\affiliation[c]{organization={Sakarya University},
	addressline={Faculty of Computer and IT Engineering, Dept. of Computer Eng.},
	city={Serdivan},
	postcode={54050},
	state={Sakarya},
	country={Türkiye}}
\ead{umit@sakarya.edu.tr}

\begin{abstract}
In web analytics, cloud-based solutions have limitations in data ownership and privacy, whereas client-side user tracking tools face challenges such as data accuracy and a lack of server-side metrics. This paper presents CAWAL (Combined Analytics and Web Application Log) as an alternative model and an on-premises framework, offering web analytics with application logging integration. This unique synthesis enables precise data collection and cross-domain tracking in web farms while complying with data ownership and privacy regulations. The framework also improves software diagnostics and troubleshooting by incorporating application-specific data into analytical processes. Integrated into an enterprise-grade web application, CAWAL has demonstrated superior performance, achieving about 24\% lower response time than Open Web Analytics (OWA) and 85\% than Matomo. The empirical evaluation demonstrates that the framework eliminates certain limitations in existing tools and provides a robust data infrastructure for enhanced web analytics.
\end{abstract}

\begin{graphicalabstract}
\includegraphics[scale=1.00]{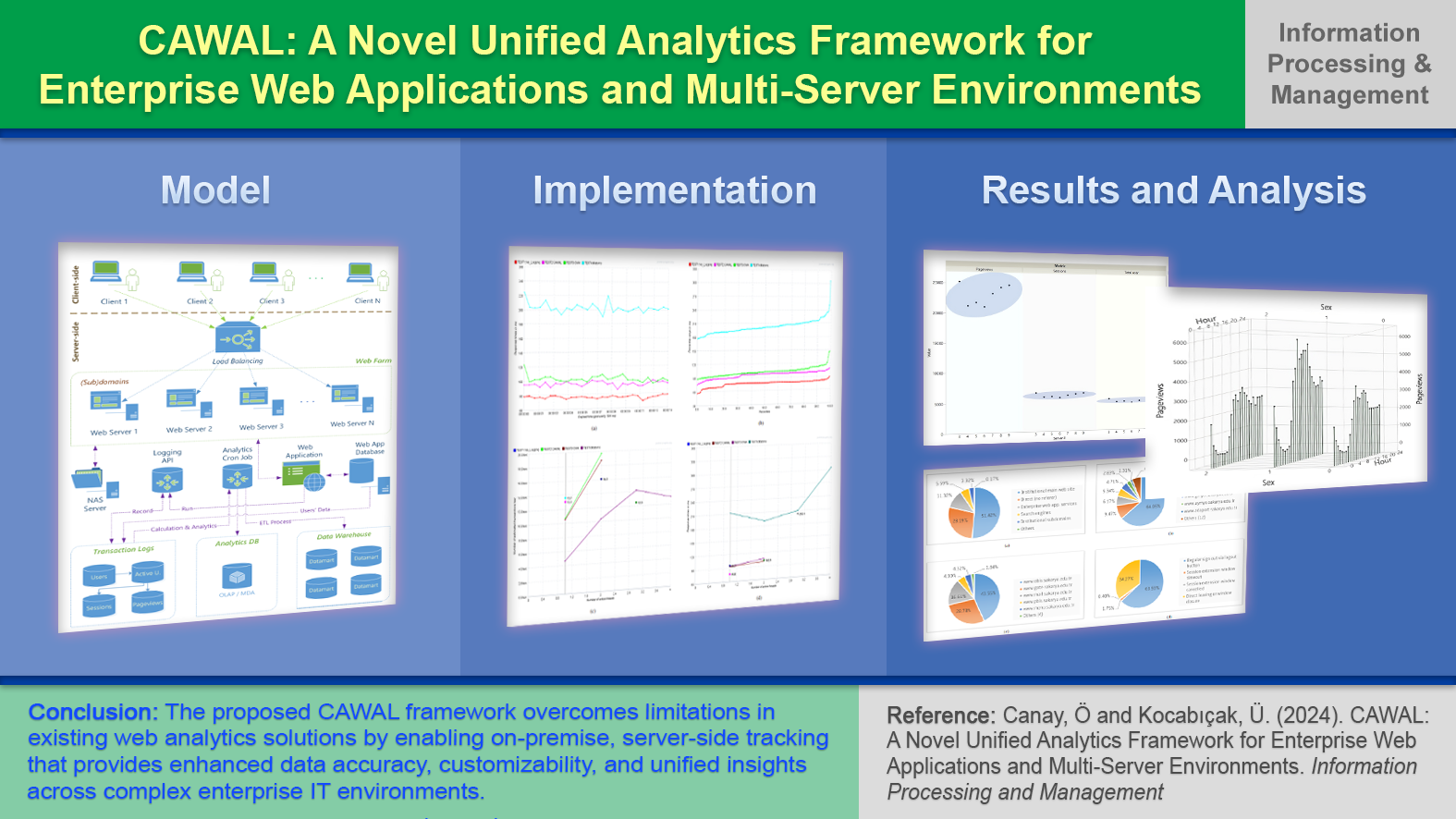}
\end{graphicalabstract}

\begin{highlights}
	\item CAWAL provides on-premises analytics with data governance.
	\item Merges web analytics and application logs in web farms.
	\item Covers SPAs, PWAs, mobile, and IoT user interactions.
	\item Boosts troubleshooting via application-centric analytics.
	\item Outperforms OWA by 24\%, Matomo by 85\%.
\end{highlights}


\end{frontmatter}


\section{Introduction}\label{sec:1}

Web analytics tools enable detailed user behavior analysis, tracking user paths, time spent on individual pages, and interactions with site elements, allowing organizations to optimize their content and design \citep{Pilz2020}. With personalization gaining importance, businesses can now segment audiences based on various parameters such as demographics, geographical location, behavior, and devices used \citep{Vamsee2023}. Furthermore, machine learning and artificial intelligence integration have enabled predictive analytics, helping to anticipate future user behavior and strategize more effectively. Early web analytics implementations predominantly relied on web server logs to capture incoming HTTP requests \citep{Kousik2021}. These logs included click streams for web traffic analysis as well as potentially misleading elements such as automatic page reloads and source files. Although, the richer the data, the more complex it becomes to analyze. Solving such complexities requires robust tools to process and visualize data, removing the challenges of detailed logs \citep{Walny2020}. Web log analyzers process clickstreams to offer key metrics on user behavior and site performance, utilizing arithmetic and statistical methods to enable precise analytics for informed decision-making \citep{Sakas2021}.

Contemporary web analytics tools present a range of capabilities and constraints in raw data access and extraction \citep{Kumar2020}. Cloud-based platforms, including Google Analytics 360 and Adobe Analytics, offer various levels of data access with features such as BigQuery Export. However, these often come with additional costs or limitations due to privacy regulations \citep{Alby2023}. Similarly, other cloud-based tools do not provide raw data access or allow limited access. In contrast, on-premises open-source tools such as OWA and Matomo offer a more flexible approach that provides comprehensive control over data. Organizations should carefully examine analytics platform data policies to meet their requirements for raw data access and export \citep{Kumar2020}.

Analytics tools relying primarily on client-side JavaScript tracking offer an incomplete view as they only capture browser-level user interactions. The evolution of Single Page Applications (SPAs), Progressive Web Apps (PWAs), mobile, Internet of Things (IoT), and ad blocker applications has only intensified these limitations, necessitating innovative solutions for effective web analytics \citep{BiornHansen2017}. However, utilizing dynamic IPs introduces additional complexities in tracking individual user behavior across sessions \citep{Canay2023}. Furthermore, multi-server architectures, essential for performance scalability in high-traffic scenarios, amplify challenges in web analytics. Uniform session management across servers in a web farm requires specialized configurations to enable cross-domain user tracking and data consistency. In addition to technical considerations, the ethical and legal dimensions of tracking have gained critical importance recently. Understanding web analytics' legal and ethical framework is crucial, especially for designing mechanisms for user consent, data transparency, and opt-out options \citep{Utz2023}. In this context, adherence to regulatory frameworks like GDPR, CCPA, and PIPEDA becomes imperative to enhance data governance and ensure ethical compliance \citep{Regueiro2021}. These regulations require robust security measures for data integrity and defense against vulnerabilities. Analytical design choices and factors like ad blockers contribute to data inconsistencies, and the need for real-time analytics introduces computational challenges. Thus, a web analytics system must balance technological, ethical, and legal aspects to be effective \citep{Demirkan2013}.

The advantages and disadvantages of web analytics tools are directly related to their characteristics. Generic web analytics tools like Google Analytics (freemium), Open Web Analytics (OWA), and Matomo (formerly Piwik) focus on user interactions on websites, capturing metrics such as page views and events but bypassing server-side application logs \citep{Alby2023}. These application logs, rich in details about software operations and errors, aren't a primary feature of these platforms, even if some integrations may allow limited log analysis. Although web analytics tools are generally user-friendly and versatile, they may not fully address an organization's unique requirements for log management. Web analytics tools and methodologies face several limitations, such as the inability to fully capture user interactions and challenges with collecting data, especially in complex web applications and multi-server environments. In contrast, a custom-built logging framework, integrated directly into the corporate web application, is tailored to the organization's unique needs, ensuring a holistic and efficient log management and analysis experience. In our prior work \citep{Canay2023}, we introduced an innovative data collection method that sidesteps the preprocessing stage in web usage mining (WUM). We build upon this foundation, providing a model and framework in this paper to fill existing industry gaps by combining application logging and web analytics, at enterprise-grade.

\subsection{Research Objectives}\label{subsec:1.1}

The primary objective of this research is to introduce and validate the CAWAL model, which integrates web analytics with application logging to provide comprehensive data management. Subsidiary objectives include:

\begin{itemize}
	\item Present the CAWAL framework as an alternative to existing tools, emphasizing its enterprise-grade capabilities and on-premises data ownership.
	\item Demonstrate CAWAL's proficiency in accurately capturing a wide range of user interactions, including SPAs, PWAs, and IoT platforms.
	\item Highlight the framework's robust session management and cross-(sub)domain user tracking in enterprise-grade web farms.
	\item Conduct a comparative evaluation of CAWAL against well-known open-source tools, Open Web Analytics (OWA) and Matomo, regarding features and performance.
\end{itemize}

The multidimensional hypothesis of this research suggests that the CAWAL framework can improve abilities such as adaptation to multi-server environments, collection of application usage data, and troubleshooting with data accuracy, diversity, ownership, and governance on an enterprise-wide scale compared to existing analytics tools.

\subsection{Primary Contributions}\label{subsec:1.2}

The main contributions of the study are:

\begin{enumerate}
	\item Introduction of the CAWAL model, emphasizing its role in bridging the gap between web analytics and application logging.
	\item Presentation of the CAWAL framework as an enterprise-grade alternative, championing centralized data ownership and privacy.
	\item Highlighting CAWAL's capacity to capture diverse user interactions, offering an on-premises data ownership advantage.
	\item Detailing the framework's proficient session management, particularly its ability to effectively track users across various subdomains.
	\item Empirical assessment that underscores CAWAL's superior performance against recognized open-source counterparts in terms of features and response times.
\end{enumerate}

\subsection{Structure of the article}\label{subsec:1.3}

\cref{sec:2} provides an overview of current web analytics solutions and highlights studies that illuminate their limitations. \cref{sec:3} delves into the design with features of the proposed model and the framework. \cref{sec:4} elaborates on the framework's application to an enterprise web information system, and its comparison with other on-premises tools is made. \cref{sec:5} presents an analysis of some results derived from applying the framework to the genuine system. \cref{sec:6} engages in a comprehensive discussion of the findings. Finally, \cref{sec:7} concludes the article by summarizing the significant insights and suggesting potential directions for future research.

\section{Related work}\label{sec:2}

Web analytics represents a complex, multidimensional domain that integrates the disciplines of data interpretation, technological application, and user behavioral studies \citep{Fundingsland2022}. Although its technical background is a computer science subject, web analytics, whose usage areas are constantly expanding, is predominantly studied in business, marketing, social sciences, health, finance, tourism, logistics, and many other fields \citep{Ahadi2022,Amalina2020}. From the computer science perspective, web analytics is closely related to disciplines such as data mining, machine learning, and artificial intelligence. 

Recent studies in the field of web analytics have addressed a diverse range of topics and methodologies. A survey conducted by \citep{Alby2023} aimed to understand the significant influence of Google Analytics on the web analytics sector. Web metrics were employed in conjunction with Google Analytics to assess the usability of e-commerce websites \citep{Kumar2022}. In-depth interviews were utilized alongside website analytics to comprehend user behavior on government websites \citep{Cheng2022}. The software tool "Web Analyzer" was introduced to analyze user behavior, relying on GET/POST methods and performance monitoring \citep{Hasan2022}. A mixed-method framework was developed to explore citizen science \citep{DeGroot2022}. Two prevalent analytics approaches, Google Analytics and Similarweb, were compared to evaluate website interactions \citep{Jansen2022}. An efficient algorithm focused on the preprocessing of weblog data was proposed \citep{Jain2021}. The application of open-source software in web analytics was examined through a case study \citep{Gamalielsson2021}. The adoption of web analytics tools by European Destination Management Organizations was analyzed \citep{Onder2021}. Web analytics tools and techniques were systematically reviewed from a business perspective \citep{Kumar2020}. In a related study, \citep{Yang2022} explored the potential of social media data analytics in business decision-making and emphasized its significance in harnessing insights from social platforms to augment business development strategies, particularly by introducing the Business Decision Making System (BDMS) framework. Matomo was compared with Google Analytics in a library setting \citep{Quintel2020}. Work was done on the identification of user behavior patterns, leveraging cross-domain log data fusion \citep{Tao2020}.

Prior work has often emphasized one or more specific areas, such as data accuracy or performance, but has not provided a comprehensive solution that addresses all these critical dimensions. Though very few examples of web analytics platform development, such as Webalyt \citep{Cegan2017}, the lack of studies describing these tools' technical infrastructure and working methods is one of the most significant gaps in computer science literature. The main reasons for this are that commercial companies dominate the field of web analytics, and it takes a very long time to develop such large projects. Computer science methods and algorithms are often used to analyze large datasets on the web, understand user behavior, and make predictions \citep{Harika2019}. Various methods of measuring website or application performance help to understand user behavior and trends, facilitating informed decisions and improvements in user experience and digital strategy \citep{Fundingsland2022}. 

\subsection{Web analytics methods and tools}\label{sec:2.1}

\begin{figure}[H]
	\centering
	\includegraphics[scale=0.55]{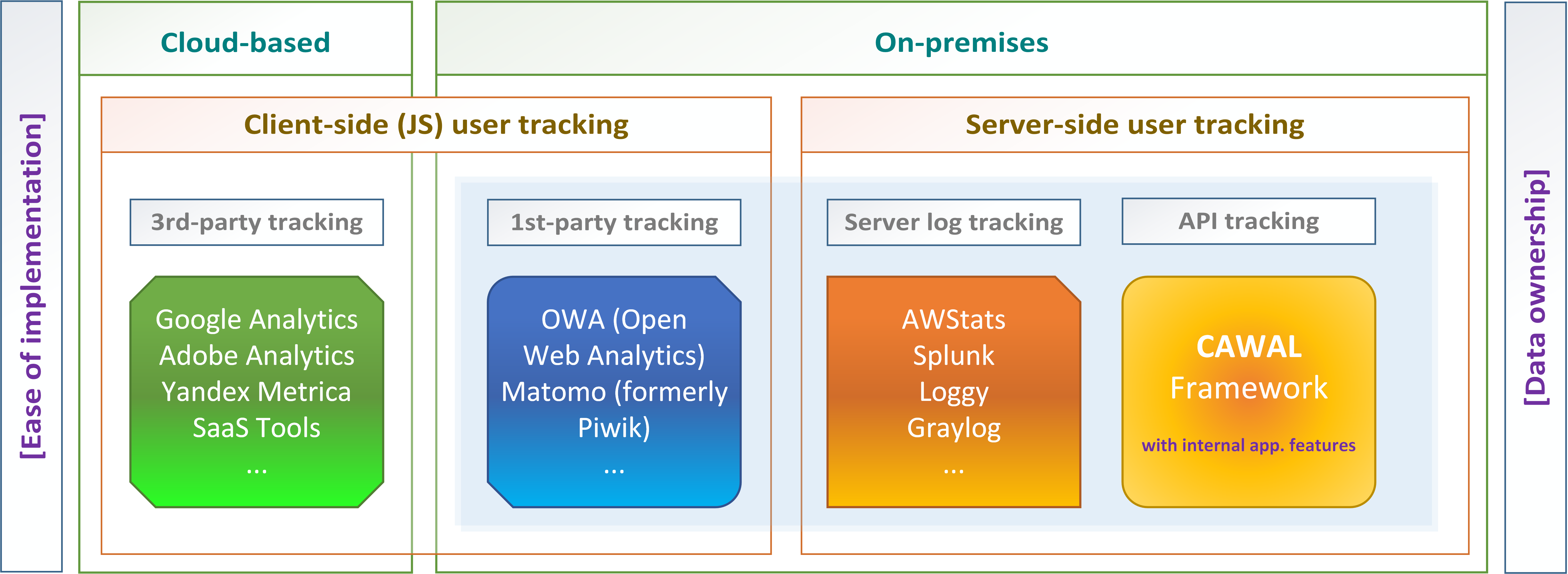}
	\caption{Commonly used data collection and user tracking methods for web analytics.}
	\label{fig:1}
\end{figure}

Web analytics are categorized into off-site and on-site types, focusing on external factors and user behaviors respectively \citep{Rafiq2022}. Modern on-site methods use server-side logging and client-side tagging, evolving from early, limited log-file-based metrics \citep{Cahyanto2022}. While advanced log analyzers such as Splunk and Graylog have emerged, traditional log-based methods have declined in popularity and application. The introduction of JavaScript tagging in the late 1990s facilitated more detailed tracking metrics, such as unique visitors and session paths, paving the way for the broad acceptance of cloud-based Software as a Service (SaaS) solutions, particularly Google Analytics \citep{Alby2023}. The methods commonly used today for data collection and user tracking in on-site web analytics and CAWAL's place among them are shown in \cref{fig:1}.

Page load speed and usability are critical for large-scale web applications, as higher latency decreases user engagement and task completion \citep{Szalek2018, Nurshuhada2019}. Studies show more than 40 percent of visitors leave the website if the page load time is over three seconds \citep{Stringam2019, Pourghassemi2019}. Precise integration of web analytics tools is essential to avoid exacerbating these issues. Design and software development teams, who are aware of this fact, pay special attention to improving the performance of web applications. Various tools and selection strategies exist in the literature \citep{Almatrafi2023,Boufenneche2022,Kumar2020}. Selecting a web analytics tool involves multiple factors, including data collection methods, location, and cost structure. Additional considerations encompass tracking capacity, real-time data provision, and designated metric monitoring. The tool's data segmentation capabilities, dashboard customization, interoperability, and cost-effectiveness also influence the decision \citep{AlOtaibi2018}. Hence, the selection should be aligned with organizational needs and web application goals without compromising user experience or performance metrics.

Free services like the basic version of Google Analytics offer initial cost-saving benefits and fundamental insights but may lack the comprehensive analytics required by organizations with high web traffic \citep{Haaksma2018,Rafiq2022}. On the other hand, on-premises platforms such as Open Web Analytics (OWA) and Matomo, which use client-side JavaScript for user tracking like their cloud-based competitors, offer alternatives focused on data ownership \citep{Alby2023}. Therefore, the choice of a web analytics tool should align not only with the organization's specific needs and goals but also with considerations such as data ownership, competitive benchmarking, and advanced reporting. This alignment is crucial for organizations that require precise and extensive analytics for decision-making.

\subsection{Data issues in web analytics}\label{sec:2.2}

Accuracy and completeness are critical for making meaningful conclusions and data-driven decisions based on data in web analytics. In a recent study focused on assessing the accuracy and integrity of web analytics data, information on measuring user interactions with websites was presented, and two analytics approaches, industry-standard, were compared \citep{Jansen2022}. Besides data accuracy, the ability to analyze data in real-time represents a significant improvement over previous static log file-based web analytics \citep{Boufenneche2022}. Current tools provide instant insights, helping to respond quickly to changing user behavior or market conditions. However, JavaScript-related problems can undermine data integrity, while multi-server and multi-domain web systems introduce unique data harmonization and analysis challenges \citep{BiornHansen2017}.

Challenges in tracking users across domains and subdomains within a web farm are exacerbated by decentralized server architectures and domain restrictions mandated by web protocols \citep{Krishnan2020}. Each server in the farm, denoted as the \textit{i${}^{th}$} server, handles a quantifiable subset \textit{n${}_{i}$} of incoming requests, logging various metrics for analysis. When \textit{m} servers exist in the system, the naive approach to data aggregation would involve summation across subsets as $\sum^m_{i=1}{n_i}$. However, this approach may overlook complexities arising from user sessions fragmented across multiple servers. Data aggregation and normalization steps become pivotal, where the objective is to merge logs from all \textit{m} servers with a time complexity of \textit{O}(\textit{m}$\mathrm{\times}$\textit{n}) in the worst-case scenario. Load balancing mechanisms can disrupt user tracking, as multiple servers may process user interactions, necessitating a unified server-side tool for data synchronization. Another imperative task is session reconciliation, which requires identifying fragmented sessions across numerous servers originating from a single user \citep{Mortazavi2020}. Such identification typically involves the utilization of session identifiers stored in client-side mechanisms like cookies. While cookies are integral for tracking user activities, their utility is confined by domain-specific limitations. Setting the cookie domain attribute to the primary domain for tracking within subdomains can ensure consistent tracking. However, tracking across distinct domains introduces heightened complexity. The diminishing effectiveness of third-party cookies, attributable to modern privacy safeguards that frequently block them, has led to exploring alternative tracking methods \citep{Durantez2023}. 

In web analytics, session, and user identification are pivotal elements for understanding user behaviors. Session identification \citep{Maslennikova2022} can be formally modeled as a function \textit{S}(\textit{A},\textit{T},\textit{C}). \textit{A} denotes user actions, \textit{T} time intervals between activities, and \textit{C} contextual data such as IP or browser. The primary objective is to formulate a function \textit{g} such that \textit{g}:\textit{A}$\mathrm{\times}$\textit{T}$\mathrm{\times}$\textit{C}$\mathrm{\to}$\textit{S}. This function aims to collate actions \textit{A} into cohesive sessions \textit{S} based on their proximity in time \textit{T} and context \textit{C}. However, achieving this involves navigating a host of complicating factors. For instance, Session Timeout (\textit{To}{}) is a pre-defined interval; if \textit{T} exceeds \textit{To}{}, it is presumed that the prior session has terminated. Similarly, periods of User Inactivity (\textit{Ia}{}) or significant Context Change (\textit{Cc}{}) could indicate the boundaries of a session. The second crucial task, user identification \citep{Salminen2020}, aims to recognize users over multiple interactions and can be represented by \textit{I}(\textit{U},\textit{M},\textit{D}). \textit{U} is the set of all users, \textit{M} stands for the methods of interaction, which can include various devices or browsers, and \textit{D} includes available data points such as IP addresses or cookies. The task function \textit{f} defined as \textit{f}:\textit{M}$\mathrm{\times}$\textit{D}$\mathrm{\to}$\textit{U} attempts to uniquely identify a user \textit{U} based on \textit{M} interaction methods and \textit{D} data points. However, several factors complicate this task. Dynamic IP allocation (\textit{Id}{}), cookies deletion (\textit{Cd}{}), browser fingerprinting ambiguities (\textit{Bf}{}), and access from multiple devices (\textit{Dm}{}) introduce uncertainties that \textit{f} must effectively navigate. Thus, both \textit{g} and \textit{f} embody the complexities and challenges inherent in these tasks.

The session boundary problem significantly impacts the accuracy of web analytics, particularly user interaction metrics. Traditional techniques employ a fixed 30-minute idle time to segment user sessions \citep{Jansen2022}, an approach that can misclassify activity. For example, reading an article without additional server requests during the idle time could be deemed an inactive session \citep{Maslennikova2022}. Such inaccuracies distort key performance indicators and introduce noise into analytics data, undermining decision-making and user experience optimization strategies.  Probabilistic models and machine learning are leveraged to improve session identification accuracy by accounting for variables like user behavior and dwell time \citep{Alhlou2016}. Another notable issue is the midnight boundary problem, which misclassifies sessions crossing midnight as multiple sessions, affecting metrics like session counts and average site time. While some established tools offer partial solutions, such as adaptive session timeout, the issue persists due to complex user behavior across time zones \citep{Jansen2022}. Therefore, both session boundary and midnight boundary problems for reliable web analytics are still significant challenges that need to be improved.

Web analytics is also considered the first stage of business intelligence development \citep{Krol2020}. Still, for some reason, cloud-based web analytics tools may not be the perfect solution for every organization. Data privacy and sovereignty concerns are at the forefront of these reasons \citep{Samarasinghe2022}. Privacy issues have become increasingly sensitive and essential in today's information societies. Countries and organizations tend to be hesitant to share their control and ownership roles over user data, especially with companies in different countries \citep{EverestPhillips2019}. While data privacy laws and regulations are vital to protecting organizations and individuals from data breaches, ensuring compliance often requires significant corporate expertise \citep{Binns2017}. In the study exploring the potential use of Matomo, an open-source software application for web analytics for academic libraries, researchers highlighted user privacy concerns with traditional tools such as Google Analytics and raised the possibility of using open-source alternatives to mitigate these issues \citep{Quintel2020}. Complementing this perspective, researchers discussed the application of open-source software such as Matomo for web analytics in the context of open government in another study \citep{Gamalielsson2021}. They emphasized that data privacy is paramount and highlighted the potential of open-source solutions to avoid vendor lock-in and achieve sustainable analytics eventually.

\section{Methodology}\label{sec:3}

Web developers typically log user actions in their applications for various straightforward purposes. Building upon this conventional practice, we have designed a novel model and framework, CAWAL (Combined Analytics and Web Application Log), to offer integrated web analytics and application logging functionalities. The model is based on the data collection approach proposed in our previous work \citep{Canay2023} and addresses the issue from a much broader perspective in the context of enterprise web systems. The framework developed as a practical implementation of the model includes an API for collecting and appropriately storing data, a database model for storing data, and a method for generating analytical information.
	
CAWAL is not a released and marketed tool but a model applied to a genuine environment. Therefore, instead of calling it a tool or a platform, describing it as a model and a framework in a broader conceptual context would be more accurate. Within the model context, defined as an abstraction representing a specific process or phenomenon, CAWAL establishes the structure that includes data collection, storage, analysis, and reporting procedures for web analytics and log management. CAWAL is also the name of the framework developed to empirically test our research's hypothesis empirically, demonstrating the model's advantages over existing web analytics tools in critical areas such as data accuracy, data ownership and governance, efficiency, and multi-server compatibility. The framework implements the conceptual model through a combination of data collection API, database structure, analytics generation, and auxiliary components tailored for enterprise-grade web analytics operations.

\subsection{Design of the model}\label{sec:3.1}

\begin{figure}[H]
	\centering
	\includegraphics[scale=0.7]{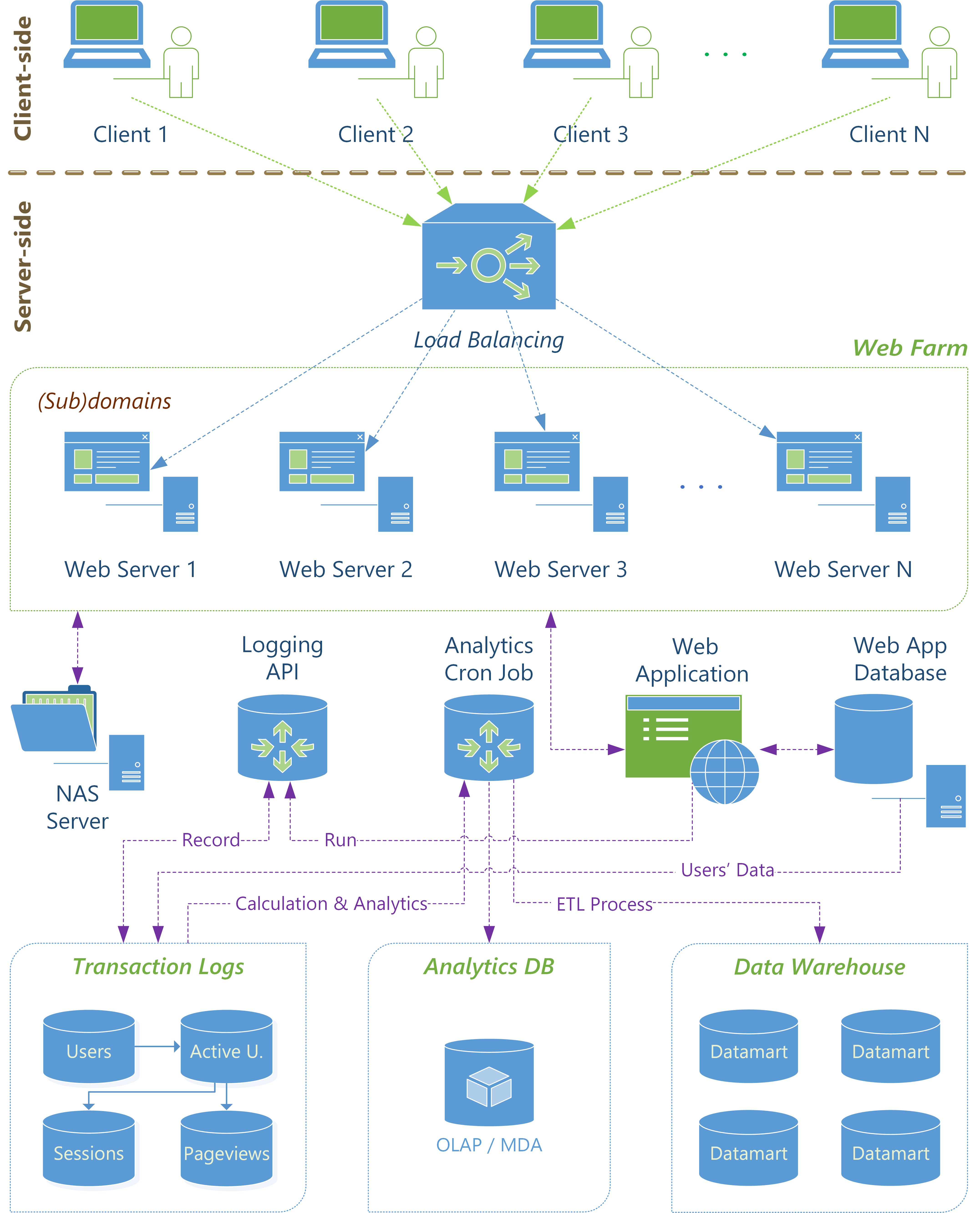}
	\caption{Full model of multi-server supported web analytics framework architecture.}
	\label{fig:2}
\end{figure}

CAWAL is designed to serve as a comprehensive architecture for multi-server-supported web analytics. Structured to operate server-side, the model supports a load balancer in the server domain to distribute incoming requests across a web farm comprising various sub-domains. \cref{fig:2} provides a detailed representation of the architecture, delineating its distinct layers and functionalities. Following the conceptual design of CAWAL, a sophisticated framework to run on Linux servers was developed to make the model viable. The architecture takes advantage of the robust and secure environment of Linux to achieve improved performance and scalability. 

According to the architecture, in a multi-server enterprise web environment, requests from clients are evenly distributed by a load balancer among the web farm's appropriate servers. These servers can access shared configuration files and session directories hosted on a NAS (Network-Attached Storage) server. As the web application responds to these requests, the logging API embedded in the web application inserts these interactions to log tables concurrently. The analytics extractor, triggered nightly by a cron job, processes this data in a dedicated analytics database. The framework also incorporates an ETL (Extract-Transform-Load) process \citep{AbdAlRahman2023} following data analytics, depositing the results in data marts within a data warehouse for subsequent analytical requirements. By integrating an OLTP (Online Transaction Processing) \citep{Li2022} database for real-time user data processing and a data warehouse for cumulative data storage, the CAWAL framework offers a scalable platform for complex web analytics operations. This model merges immediate data collection with long-term analytical storage, creating an adaptable and scalable solution for contemporary web analytics challenges.

The CAWAL framework uses object-oriented PHP and integrates closely with a MySQL relational database, facilitating a robust analytics solution. PHP's functionalities, including session management, form the backbone of server-side operations, while MySQL serves as the primary database system for efficient data storage and query execution \citep{Alya2022}. Specific optimizations in MySQL indexing are designed for high-speed data retrieval, an essential feature for real-time analytics. Additionally, the framework utilizes MySQL's ACID (Atomicity, Consistency, Isolation, and Durability) compliant transactional capabilities for reliable data persistence and leverages built-in SQL functions for data processing, statistics, and time series analysis. The architecture also accommodates distributed capture agents through CAWAL's API, resulting in a comprehensive web analytics solution that balances scalability, reliability, and performance.

\subsection{Data model}\label{sec:3.2}

\begin{figure}[H]
	\centering
	\includegraphics[scale=0.203]{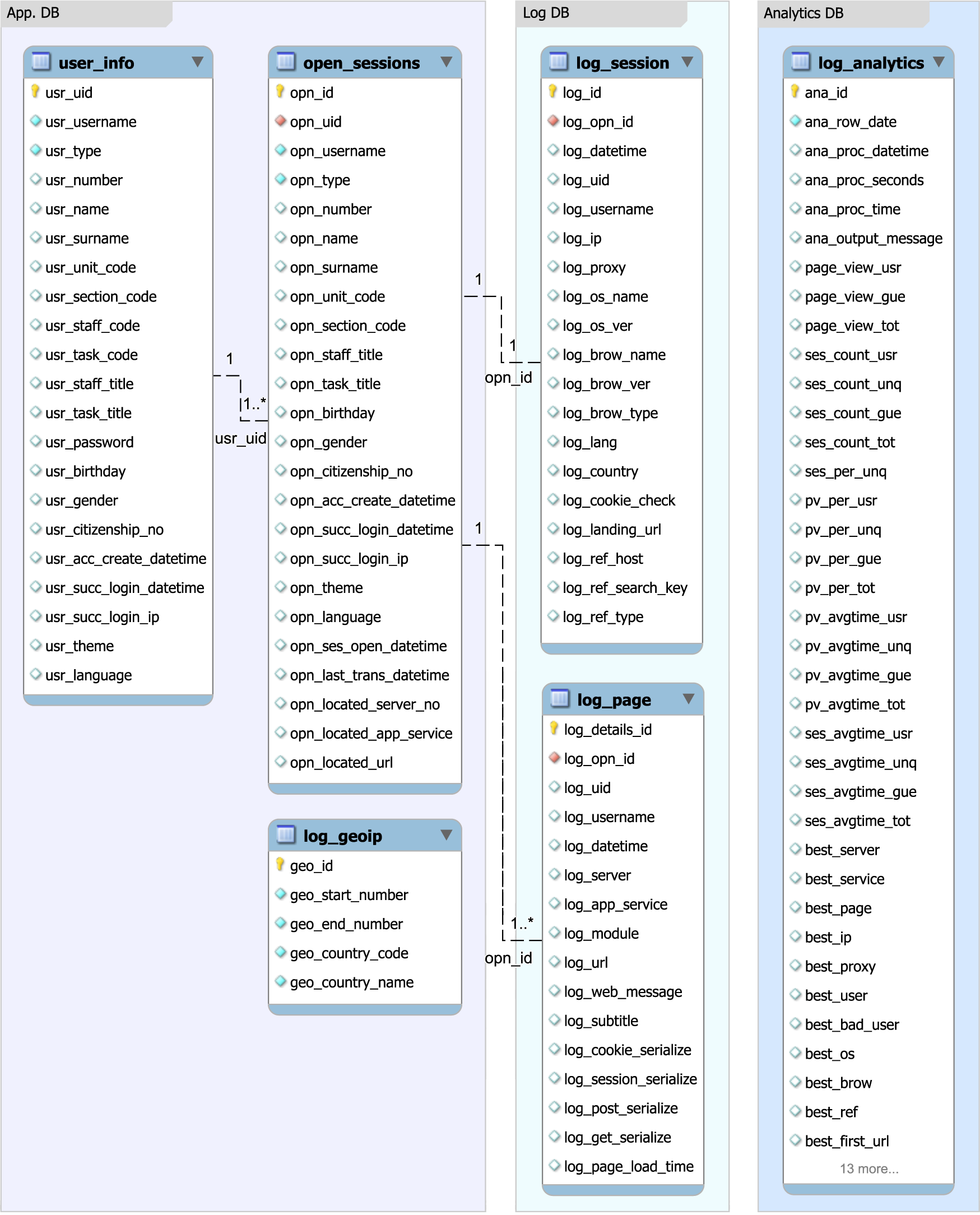}
	\caption{UML diagram of the logical data model.}
	\label{fig:3}
\end{figure}

CAWAL implements an optimized lean data model to enable efficient analytics while minimizing storage overhead. The framework designed in accordance with the model incorporates an Application Programming Interface (API) \citep{Ofoeda2019} for the structured acquisition and storage of data, along with a database schema for data retention. The data collection infrastructure underpinning the framework and the two main database tables that hold the usage data, "log\_session" and "log\_page", are described in detail in our previous paper \citep{Canay2023}. The UML diagram in \cref{fig:3} provides a comprehensive visualization of the logical data model, including the analytical database, which was not included in our previous work.

The "log\_session" table, one of the two main database tables of the CAWAL framework, is a critical component in understanding and monitoring user interactions within a system. Designed to represent each user session as a distinct row, this table houses extensive information concerning the user and the technical environment. As a key to web analytics, the "log\_session" table includes details such as session ID, user ID, IP address, and proxy address. The table also encapsulates information such as the operating system and its version, browser name, version, type, language, country (via the GeoIP table), the initial URL accessed within the site, referring site, search engine keywords, and processing time. Together with the meticulous recording of various parameters, this table serves multiple purposes. \cref{tab:2} shows a sample tuple from the ``log\_session'' table, reflecting the comprehensive approach the CAWAL framework employs to gather data related to user sessions. The table aids in detecting patterns and trends in user behaviors, understanding users' demographic distribution, and obtaining technical insights, such as browser preferences and operating system utilization. The ability to trace the initial URL and referring sites adds another layer of complexity, allowing for a more profound understanding of user navigation patterns and source preferences.

\begin{table}[htbp]
	\centering
	\begin{threeparttable}
	\caption{A sample tuple from the ``log\_session'' table.}
	\label{tab:2}
	\small
	\begin{tabular}{ l l l }
		\toprule
		Column & Source & Sample Data\\
		\midrule
		log\_id & Database & 3242335\\
		log\_opn\_id & Application & 60915392\\
		log\_datetime & Web server & 10.05.2020 22:42\\
		log\_uid & Database & 540\\
		log\_username & Database & canay\\
		log\_ip & HTTP & 193.140.253.81\\
		log\_proxy & HTTP & \\
		log\_os\_name & User-agent & Windows\\
		log\_os\_ver & User-agent & 11\\
		log\_brow\_name & User-agent & Chrome\\
		log\_brow\_ver & User-agent & 114\\
		log\_brow\_type & User-agent & 1\\
		log\_lang & HTTP & tr\\
		log\_country & GeoIP DB & Turkiye\\
		log\_cookie\_check & HTTP & True\\
		log\_landing\_url & Web server & https://www.gate.sakarya.edu.tr/\\
		log\_ref & HTTP & Google\\
		log\_ref\_host & HTTP & www.google.com.tr\\
		log\_ref\_search\_key & HTTP & sau gate\\
		log\_ref\_type & HTTP & 4\\
		\bottomrule
	\end{tabular}
	\end{threeparttable}
\end{table}

Another core database table of the framework, ``log\_page'', represents a comprehensive mechanism for keeping track of transactions between pages while also containing application-specific data such as application header, application message, cookie, session, post, get, and page load time. The table, specifically designed to enhance the framework's functionality, aids in software quality improvement, facilitating error detection processes and ensuring robust application performance by providing detailed information on various aspects of the application. Moreover, the web administrator can leverage these database tables to monitor system usage and gain insights through real-time analytics information. A notable feature within this framework is the emphasis on obtaining page production times, a critical factor for web developers. This information improves web performance optimization by enabling them to detect database or code-induced slowness, reinforcing the CAWAL framework's commitment to providing a highly efficient and accurate analytics solution. The "log\_analytics" table houses complex analytics information processed for one day in each tuple. The table contains about 50 fields, the last six of which are serialized multidimensional data. The one-day record in the table is about 68 kilobytes, which indicates the scope of the analytics extraction.

\subsection{Data collection}\label{sec:3.3}

\begin{figure}[H]
	\centering
	\includegraphics[scale=0.72]{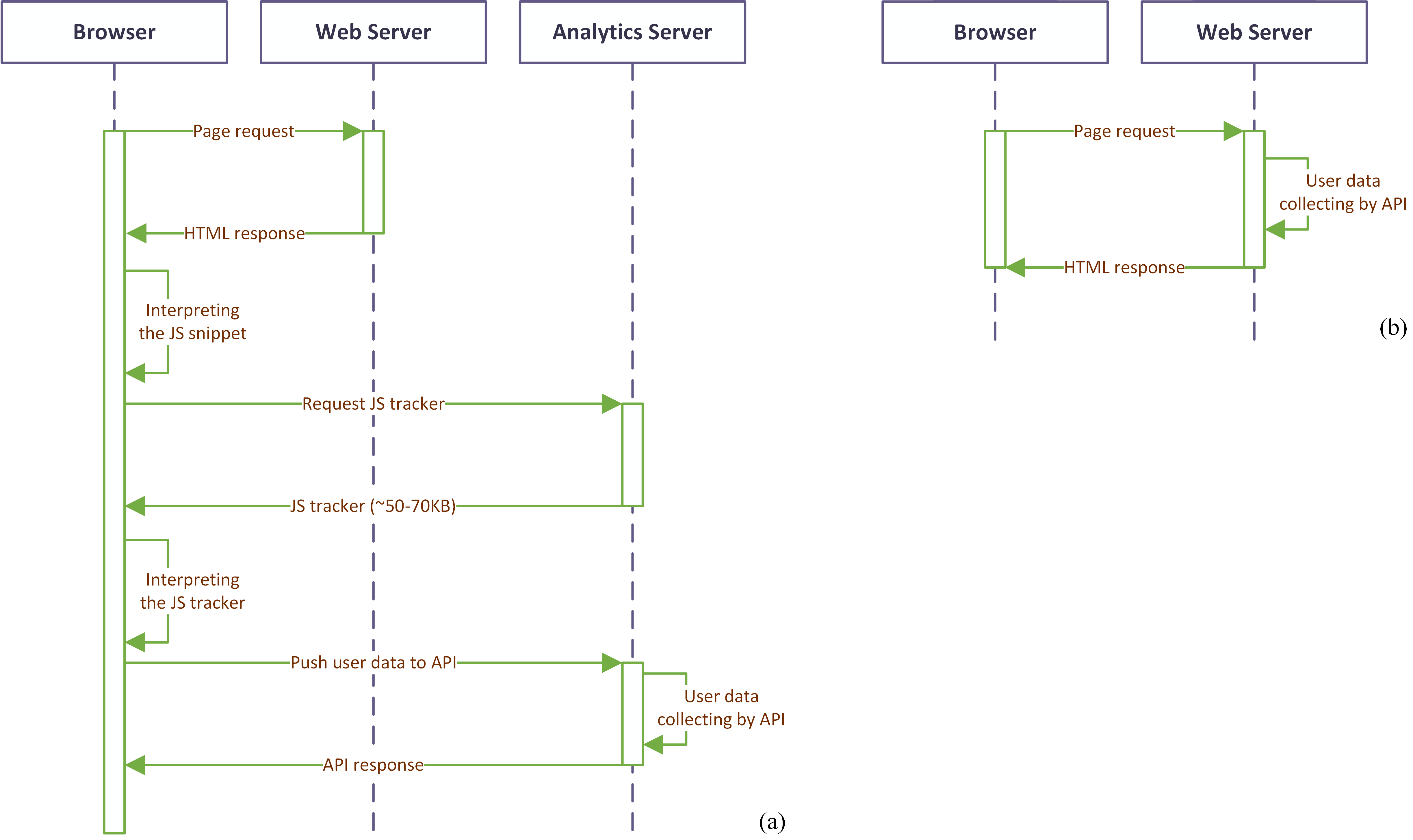}
	\caption{Sequence diagram of client-side (a) and server-side (b) user tracking.}
	\label{fig:4}
\end{figure}

Metrics and dimensional data, the core components of the data collection model, are obtained from four different sources:  HTTP request, network-level, application-level, and external data \citep{Canay2023}. HTTP request data is captured from HTTP request messages through HTTP headers. Though server-generated and associated with HTTP requests, network-level data is not part of the HTTP request but is necessary for successful transmissions (e.g., client IP address). Application-level data, transmitted with HTTP requests, includes session and referral information managed by server-side web programming languages. External data, typically in the web application database, can be associated with the other categories and includes elements like country identification derived from the IP address and user profile information. These various sources are then processed and transformed into metrics and other valuable data. In web analytics, data collection combines data from various sources through user tracking and data fusion. The CAWAL framework collects and processes such information from multiple sources through a data collection API that works server-side and application-based to acquire, interpret, format, and store data. As shown in \cref{fig:4}, third-party tools send two extra requests to the log server for user tracking and data collection, while CAWAL interacts with the server much less than others due to its lean nature. 

The data collection process is rooted in collecting and analyzing both qualitative and quantitative information. Qualitative attributes, or dimensions, include aspects such as the client's browser and operating system, the IP address, preceding webpages, channel identification, geolocation, and specific user profiles. These data can be obtained through diverse sources such as HTTP requests, network protocols, applications, or external means. Conversely, quantitative measures or metrics include pageviews and application-specific data obtained from HTTP requests or external sources. Incorporating these dimensions and metrics into the CAWAL framework, after undergoing various cleaning, subtraction, or calculation processes, provides a precise understanding of user interactions and behaviors. Detailed analysis of this data also enriches decision-making in areas such as user experience improvement, site optimization, and strategic marketing planning.

\subsection{Data processing and storage}\label{sec:3.4}

In the CAWAL framework, transaction logs are systematically aggregated during routine operations for subsequent data processing and storage. These logs are processed daily to generate valuable statistics stored in a single database record. The architecture serves various purposes: it calculates server load and user density by the hour, evaluates user navigation details like page visits and dwell times, identifies slow-loading pages for optimization, and detects unauthorized access attempts. Additionally, it logs key messages and titles from the page and captures data from cookies, sessions, and HTTP methods such as POST and GET. This organized data serves not only as the foundation for generating comprehensive reports and statistics but also for enhancing system performance and security measures. The subsequent sections outline the steps involved in handling these transaction logs.

\textbf{Data backlog.} Especially in large-scale systems, the number of records generated in daily transactions can exceed millions. The sustainability of a database table of this scale in an OLTP structure, its usability for analysis purposes, and the manageability of backup processes will not be efficient. To solve this problem, the CAWAL model employs a comprehensive ETL process. Daily log data is transferred to temporary transaction tables at the beginning of the analytics extraction process after midnight, and OLTP tables are emptied and optimized. After the analytics extraction process on the temporary tables, the processed data is transferred to the data warehouse for long-term storage and future use. Thus, the problem of data backlog and slowdown in OLTP tables is eliminated.

\textbf{Analytics extraction process.} Analytics extraction within the CAWAL framework is an automated process carried out nightly after midnight. This process begins with database maintenance operations, followed by session and page view information processing from the preceding day. Analytical extraction steps are then executed to create daily web analytics data. This comprehensive procedure performs highly complex and sophisticated queries involving various SQL commands categorized as DDL, DML, and DQL \citep{Brahmia2022}. These queries, the most critical component of the analytics extraction phase, result in a significant amount of single or multi-dimensional data. Multi-dimensional data in the form of arrays are serialized and stored in the database. Optimizing database indexes and extensive queries is critical for data collection and analytics performance in systems that process millions of records daily. To achieve this, considerable enhancements have been made to the database and the algorithm, resulting in a highly efficient script. Critical information about the entire process is stored on an event-by-event basis using timestamping in a text-based log file without any dependency on the database. This ensures that the information concerning the script's operations is preserved reliably and can be effortlessly monitored over the web.

\textbf{ETL process and data warehouse.} CAWAL incorporates a robust ETL process to structure captured web usage data into a specialized data warehouse. This process starts with analytics extraction, aggregating data into a central data store. Subsequently, tailored transformations are executed to convert this raw data into an analytical schema conducive to reporting and mining. Post-transformation, the operational data is routed to the respective monthly data mart within the data warehouse, utilizing specialized procedures to manage transitions between months and years. This data warehouse strengthens CAWAL's analytical capabilities by providing an environment conducive to multi-dimensional analysis such as OLAP \citep{Keskin2022}. It permits the correlation of various metrics---ranging from sessions and users to pages, timestamps, and traffic sources---thereby enabling the generation of detailed long-term reports, uncovering trends and patterns, performing segmentation analyses, and applying diverse data mining algorithms.

\subsection{Features and functionality}\label{sec:3.5}

The need for robust, adaptive, and responsive tools has become paramount in contemporary web analytics, especially in large-scale and complex web environments. In addressing these demands, the CAWAL framework aims to improve efficiency, reliability, and user experience by providing features and functionalities that reflect the complexity of modern web systems. Specific features and functionalities contributing to the framework's effectiveness are detailed below.

\textbf{Multi-server environment adaptation.} The CAWAL framework is notably designed to be compatible with large-scale systems and the multi-server environment called web farm and enables distributed computing over a network of servers. This architecture not only facilitates load balancing but also ensures a high degree of scalability and reliability. When real-time processing of tracking data is required, the distributed nature of the web farm allows for rapid computations. Using this distributed structure, CAWAL balances the demand load efficiently, manages computational tasks, and potentially improves system performance. The web farm configuration is integral to CAWAL's approach to addressing the analytics process for large-scale web applications. It reflects a strategic alignment with contemporary technological demands and optimization principles.

\textbf{Cross-domain user tracking.} The CAWAL framework is engineered to manage user sessions effectively in environments that utilize multiple Apache servers. By using a common session directory located on a network-attached storage (NAS) server, CAWAL ensures data consistency for users navigating various web services. To track sessions across different domains and subdomains, the framework includes specific database fields such as "server number" to pinpoint the exact server processing each request and "service" to identify the application service corresponding to each subdomain. This methodological approach enhances session data management and broadens the analytical scope of the system, enabling comprehensive, multi-dimensional analyses of user interactions across servers, domains, and services.

\textbf{Session timeout notification.} The CAWAL framework incorporates the session timeout notification method, an essential feature for enhancing user engagement and preserving session continuity. By utilizing this method, the framework monitors user inactivity within a defined time frame and, upon approaching the timeout threshold, triggers a notification to the user. This notification informs the user that the session will expire and provides an option to continue if desired. The primary advantage of this approach within the CAWAL framework lies in its capacity to minimize inadvertent session termination, thus reducing potential data loss and improving the overall user experience. This method, mainly used in applications where sensitive information is processed, such as Internet banking, strengthens user awareness by actively involving the user in session management and helps provide a secure and controlled environment. The method further resolves the issue of session duration, a problem other tools often address by defaulting to a 30-minute assumption, thereby offering a precise definition of sessions.

\textbf{Midnight boundary problem solution.} Accurate identification of sessions that cross midnight is important for preventing misleading metrics caused by misclassification and for maintaining the integrity of web analytics. Many web analytics tools extend the 'session timeout' threshold or define it as a different session when transitioning into a new day to address this problem. However, CAWAL focuses on the start of the session rather than individual page request times. The analytics extraction script running in pre-dawn hours treats sessions that start before midnight and continue into the next day as a continuous session and includes them in the previous day's statistics. This enables a more robust and accurate assessment of user interactions, eliminating the risks of metric distortion that can lead to incorrect strategic decisions.

\textbf{Dynamic IP compliance.} In CAWAL's session tracking, sessions are initiated by the web server, and two distinct devices using the same IP address are considered separate sessions. Since it operates based on the server session variable, there is no risk of mixing sessions with identical IP addresses, as could happen with web server logs. These sessions are terminated when the user logs out or does not respond to the warning window activated when there is no page activity for a certain period. CAWAL's ability to accurately identify the client, even in the face of dynamic IP distribution, is an essential advantage in ensuring accurate tracking of users.

\textbf{Data ownership and governance.} Data ownership and governance are imperative in addressing today's fundamental data security and privacy challenges. As an on-premises solution, CAWAL enables data ownership and gives organizations complete control over their analytics data assets. Data is retained within institutional infrastructure, ensuring privacy and security based on internal policies. However, compliance with localized data regulations becomes tractable. On the other hand, data governance capabilities like access controls, audit logs, and encryption enhance trust and transparency. Institutions can tailor data retention strategies due to complete control over analytics pipelines. CAWAL facilitates organizational autonomy in web intelligence by transferring control from external vendors to internal parties, aligning practices with ethical standards and values while maintaining data ownership and utilizing in-depth analytics.

\textbf{Management interface.} The management interface of the CAWAL framework is a hub that offers advanced functionalities to command and control various aspects of the web analytics process. Having a user-friendly interface, the dashboard gives administrators easy access to critical functions such as user management, security settings, data configuration, and customization of reporting features. The dashboard's modular design offers streamlined navigation, allowing for efficient adjustments to the system's parameters and settings. Designed to respond flexibly to organizational requirements, the panel contributes to the effective handling of web analytics.

\textbf{Real-time monitoring.} For web and system administrators, monitoring the real-time status of a web system, including servers, web services, users, and sessions, is critical for rapid problem detection and resolution. This real-time analysis enables proactive responses before problems escalate, seriously impacting operations and the user experience. Thanks to the CAWAL framework, comprehensive information about the instant status of the system, such as how many visitors are on which server, service, or module, how many of them are guests or users, or how many visitors with the same IP address are among them, can be easily monitored via the admin panel. In addition, administrative operations such as banning, unbanning, kicking users within the scope of authorizations can be easily performed through the real-time user monitoring interface.

\textbf{Reporting and visualization.} The CAWAL framework provides advanced reports to present complex web analytics data in an accessible way. Tailored to extract diverse metrics such as user engagement, traffic sources, and behavioral patterns, the reporting component offers customization to align with specific business goals and strategies. Complementing this, the visualization feature transforms raw data into various interactive visual formats such as charts, graphs, or heatmaps, providing real-time insight into web events. This combination of reporting and visualization supports intuitive understanding, trend identification, and data-driven tactical and strategic decisions.

\section{Implementation and evaluation}\label{sec:4}

The designed CAWAL framework was deployed on the Campus Automation Web Information System (CAWIS)  \citep{Canay2011}, the corporate web application of Sakarya University, to substantiate the multi-dimensional hypothesis. The framework has been highly influential in monitoring the entire system, including server and web service densities, understanding user behavior and usage practices, configuring the web application accordingly, and improving application quality and security. Valuable results from the implementation affirm the framework's advantages over existing tools in various dimensions, such as performance and data governance.

\subsection{Web application integration}\label{sec:4.1}

The CAWAL framework is embedded into application code to record events like exceptions, user flows, state changes, etc., that are invisible to external trackers. APIs allow the insertion of custom app-specific data like form fields into tracking logs. Session data is supplemented by monitoring application servers to paint a complete picture. CAWAL's data collection API is integrated within the web application, specifically in the initial part of the code. This integration enables the API to be invoked with every page request, aligning with the systematic design of the application. When the application is correctly constructed and the API is seamlessly integrated, it abstracts the logging complexity, allowing application developers to focus on their core functionalities. During the writing and deployment of application codes, CAWAL's API consistently operates within the layered framework, collecting data unobtrusively. Before completing the web application code, details about the operations on the active page, such as the page loading time, database query delays, and any occurring errors, are updated in the page log table by the API. With code-level visibility, CAWAL gains unique insights that are hard to extract otherwise. This systematic approach facilitates the continuous gathering of essential data, contributing to the ongoing improvement and optimization of the application.

\subsection{Web farm integration}\label{sec:4.2}

Implementing CAWAL in a genuine corporate web application encompassing ten web servers and many subdomains used to deliver web services epitomizes a sophisticated approach to handling extensive web traffic and analytics requirements. The CAWAL has worked efficiently with a well-configured load-balancing mechanism to keep track of transactions on different servers simultaneously, ensuring that the system maintains strong performance even during periods of heavy user activity. Furthermore, using a NAS server to which the session and settings folders of the web servers have routed adds another layer of consistency to the CAWAL implementation. This centralized storage solution ensures uniform and organized session data management, guaranteeing continuity across the web farm. This architecture, enriched with strategic load balancing and NAS server utilization, demonstrates the system's ability to scale efficiently, provide uninterrupted service, and deliver a flexible solution for complex web applications.

\subsection{Time complexity of the framework}\label{sec:4.3}

Server-side application-based web analytics, structured through the CAWAL framework, begins with log collection, where each of the \textit{n} user requests to the server creates a log entry. This results in a linear time complexity, \textit{O}(\textit{n}). The following phase is data transformation, wherein each log entry might undergo \textit{k} transformations, such as converting IP addresses to geographic locations, resulting in a complexity of \textit{O}(\textit{kn}). Database storage of this processed information adds another layer of complexity, dependent on factors such as the database's structure and indexing. More complex operations are encountered in subsequent stages. Although grouping data by sessions or users during normalization leads to quadratic complexity \textit{O}(\textit{n}${}^{2}$), optimized algorithms reduce it to \textit{O}(\textit{n}log\textit{n}) or linear time. Aggregation and summary operations produce data summaries with variable complexity. Simple aggregations may be linear, while complex ones are more computationally demanding.

\subsection{Feature comparison}\label{sec:4.4}

Comparing the methods and tools used for web analytics can be made qualitatively and quantitatively. It is possible to make an empirical evaluation by examining selected methods and tools according to various criteria in terms of qualitative characteristics. A feature comparison of web server logs, Google Analytics (freemium), open-source web analytics tools (OWA and Matomo), and the CAWAL framework is given in \cref{tab:2}.

\begin{table}[H]
	\resizebox{\textwidth}{!}{%
	\centering
	\begin{threeparttable}
	\caption{Comparison of diverse methods and tools in web analytics based on features and capabilities.}
	\label{tab:2}
	\small
		\begin{tabular}{ l c c c c  }
			\toprule
			& \parbox[t]{2.5cm}{\centering Web Server Logs \\ (Apache)} & \parbox[t]{2.5cm}{\centering Google Analytics \\ (Cloud)} & \parbox[t]{2.5cm}{\centering OWA/Matomo \\ (On-premises)} & \parbox[t]{2.5cm}{\centering CAWAL \\ (On-premises)} \\
			\midrule
			Platform independence & No & Yes & No & No \\
			Application independence & Yes & Yes & Yes & No \\
			Intensive local resource utilization & No & No & Yes & Yes \\
			Setup requirements and coding effort & Low & Low & Medium & High \\
			User tracking (sided) & Server-side & Client-side & Client-side & Server-side \\
			User tracking (based) & Web server-based & JavaScript-based & JavaScript-based & Application-based \\
			Single HTTP request to web (log) server & Yes & Yes & No & Yes \\
			Additional load to the client browser & No & Yes & Yes & No \\
			Additional load to the web server & No & No & Yes & Yes \\
			Cross-(sub)domain tracking & Limited & Limited & Limited & Yes \\
			Session timeout notification & No & No & No & Yes \\
			Session boundary detection & Configurable & Improved & Improved & Adaptive \\
			Midnight boundary handling & Baseline & Partial & Partial & Full \\
			SPA and PWA integration & Natural support & With Effort & With Effort & Natural support \\
			Mobile and IoT compatibility & Natural support & With Effort & Partially & Natural support \\
			Application-specific (custom) data & No & Limited & Limited & Unlimited \\
			Local data warehouse infrastructure & No & No & No & Yes \\
			Data accumulation method & Local (Service) & Cloud (SaaS) & Local (API) & Local (API) \\
			Free access to raw data & Yes & No & Yes & Yes \\
			Data sovereignty and ownership & Yes & No & Yes & Yes \\
			Type of collected data & Semi-structured & Structured & Structured & Structured \\
			Complexity of data transformation & High & Medium & Low & Medium \\
			Suitability of data for WA & Low & High & High & High \\
			Suitability of data for WUM & Medium & No & Medium & High \\
			\bottomrule
		\end{tabular}
		\end{threeparttable}
	}
\end{table}

Although a comparison with free solutions has been made here, it is worth noting that some of the restrictions in the freemium version of Google Analytics are not available in the paid version, GA360. These methods should also be considered regarding capabilities such as performing real-time analysis, operating under heavy load, obtaining non-standard and application-based data, privacy, and data security. The proposed model is quite successful compared to traditional methods in terms of its advantages, such as data accuracy, completeness, timeliness, privacy, scalability, and ability to adapt to future data needs and insight research. CAWAL, which integrates with other tools and systems the organization uses, is designed in a structure that can handle increasing amounts of data and user traffic over time.

\subsection{Performance comparison}\label{sec:4.5}

Due to their inherent differences, comparing client-side and server-side web analytics tools and measuring their effectiveness against each other presents complex challenges and requires a combination of sophisticated approaches. In addition to the qualitative approach, the differences in the page loading times on the client side to be examined within the framework of searching for common ground between the methods can be a quantitative performance indicator. In this section, the performance of the proposed CAWAL framework is compared with the leading open-source analytics tools, OWA and Matomo. However, it is imperative to highlight the inherent limitations restricting the direct comparison of these systems with SaaS web analytics tools like Google Analytics and Adobe Analytics. To create the experimental environment, CAWAL, OWA, and Matomo frameworks have been made to run locally with the sample corporate web application on a workstation with an Intel i7 processor and 16GB of RAM. All performance measurements, including server and client roles, were performed on a Windows 11 OS workstation. For web applications to work correctly, all hostnames, including subdomains, are redirected to localhost via the "hosts" file of the Windows OS. The logging framework selection during the request has been provided with a GET parameter added to the main page of the experimental website. According to the incoming parameter value, the JavaScript snippet of the relevant framework has been included in the HTML content and sent to the client.

\subsubsection{Client-side performance analysis}\label{sec:4.5.1}

In web analytics tools that perform data gathering on the client side, JavaScript-generated session and user ID, referrer, screen resolution, and other information that needs to be transmitted are added as parameters to the GET or POST type API request. The data collector API can run on the same server as the web application or be located on a different server. However, in this experimental study, the entire environment was set up on a single computer since the load of the selected web analytics tools on both the client and the server side will be determined. The CAWAL log framework does not involve any client-side process, while two third-party requests are made, one script and one API, in both other client-side tools. Sending a request to API after downloading and interpreting a JavaScript file over 50 kilobytes to track the user causes some latency on the client side. Although trace scripts usually run asynchronously, they also load the server and the client's processor. However, as an advantage of their asynchronous execution, their negative impact on users' time-to-feel interaction (TTI) \citep{Hossfeld2018} is relatively low. 

\begin{figure}[H]
	\centering
	\includegraphics[scale=0.56]{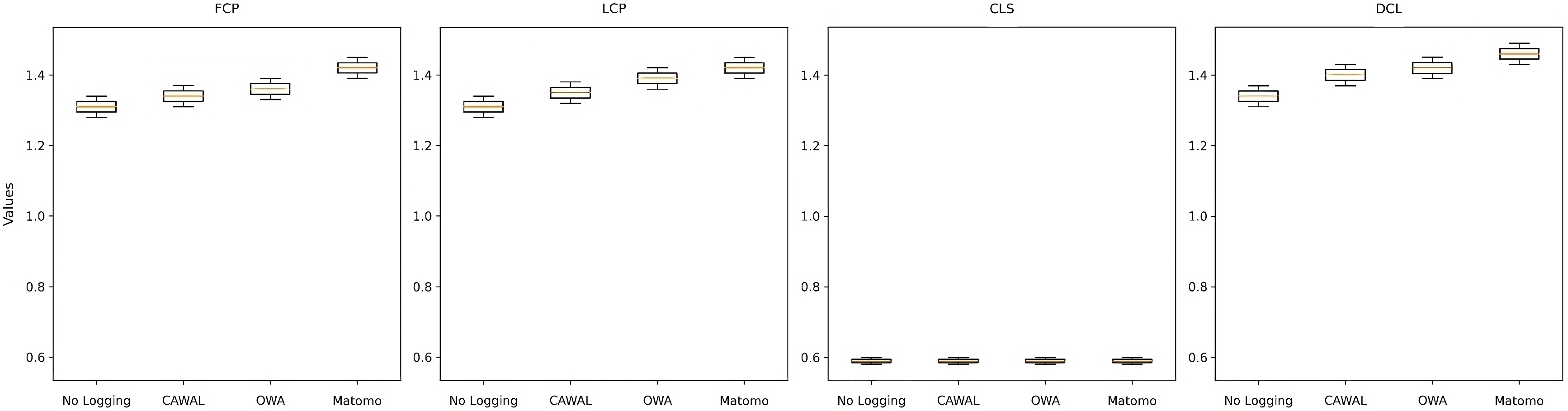}
	\caption{Comparison of equivalent on-premises web analytics tools based on client performance metrics.}
	\label{fig:5}
\end{figure}

Page loads performed in this context have been measured through the performance insights tool in Chrome DevTools \citep{Bielak2022} with fast 3G and no CPU throttling. Average measurement times were obtained in seconds by performing N=10 consecutive repetitions. Numerous studies show that evaluating results through the median of five consecutive runs significantly reduces variability and is sufficient to reduce the testing process to a reasonable time frame \citep{ChanJongChu2020,Yu2021,Hericko2021}. Testing on different devices, such as a high-performance desktop and a low-performance laptop, will fluctuate the measurements. However, measuring the performance of the frames compared to each other in the same environment and averaging these measurements by repeating them many times eliminates this effect and makes the measurements more meaningful. As the importance of user experience continues to grow in the development of web applications, a range of performance metrics have been established to assess and optimize page load speed and responsiveness accurately \citep{Armaini2022}. FCP, LCP, and DCL collectively offer a comprehensive picture of a web page's performance and user experience on the client side \citep{Dewi2023,Saif2021}. \cref{fig:5} presents the performance measures of the analytics platforms considered according to these metrics.

The measurement result shows that the tools perform close to each other with slight differences on the client side. The fastest page load on the client side was, as expected when the web application ran without any analysis tool. CAWAL, with its lean structure, made the page load faster than the other tools, followed by OWA and Matomo. The Matomo generally exhibits the highest values in the FCP, LCP, and DCL metrics, while CAWAL's performance aligns with expectations for faster page-loading capability. The CLS metric reveals minimal variation between frames, indicating uniform layout stability. The information from the controlled conditions provides valuable guidance in selecting an appropriate framework based on specific performance requirements and optimization goals.

\subsubsection{Server-side performance analysis}\label{sec:4.5.2}

\begin{figure}[H]
	\centering
	\par\medskip
	\begin{subfigure}[t]{0.495\linewidth}
		\centering
		\includegraphics[scale=1.20]{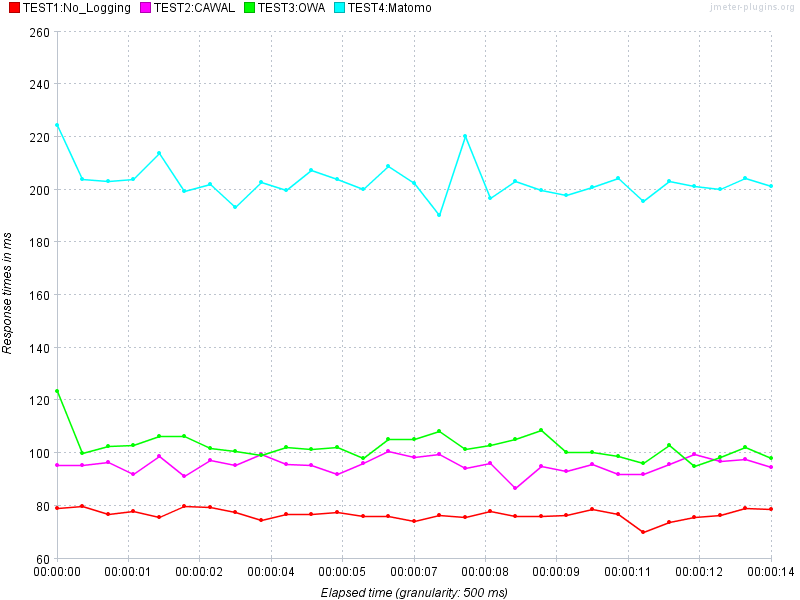}
		\subcaption{}\label{fig:6a}
	\end{subfigure}
	\begin{subfigure}[t]{0.495\linewidth}
		\centering
		\includegraphics[scale=1.20]{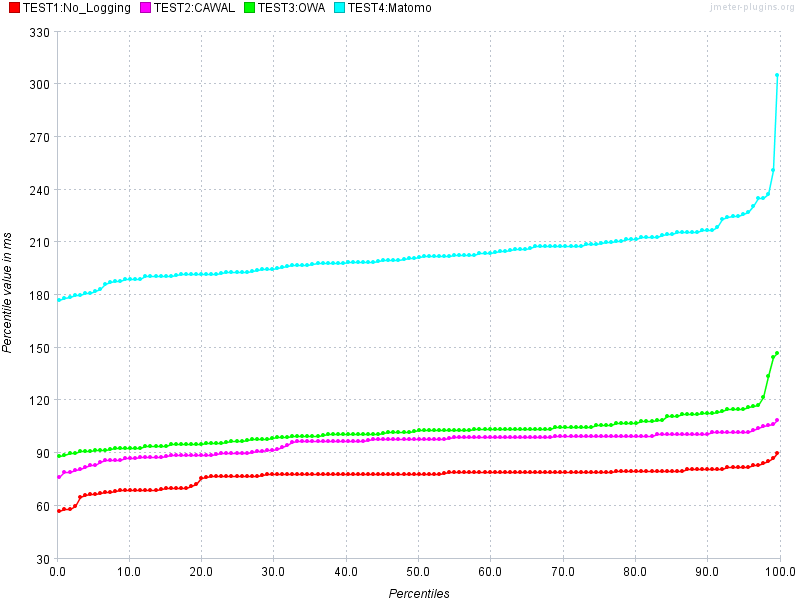}
		\subcaption{}\label{fig:6b}
	\end{subfigure}
	\par\medskip
	\begin{subfigure}[t]{0.495\linewidth}
		\centering
		\includegraphics[scale=1.20]{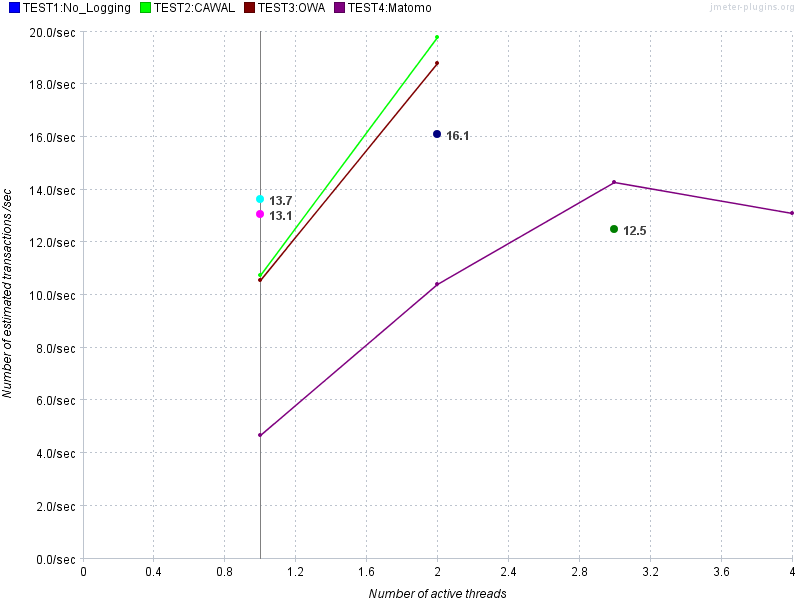}
		\subcaption{}\label{fig:6c}
	\end{subfigure}
	\begin{subfigure}[t]{0.495\linewidth}
		\centering
		\includegraphics[scale=1.20]{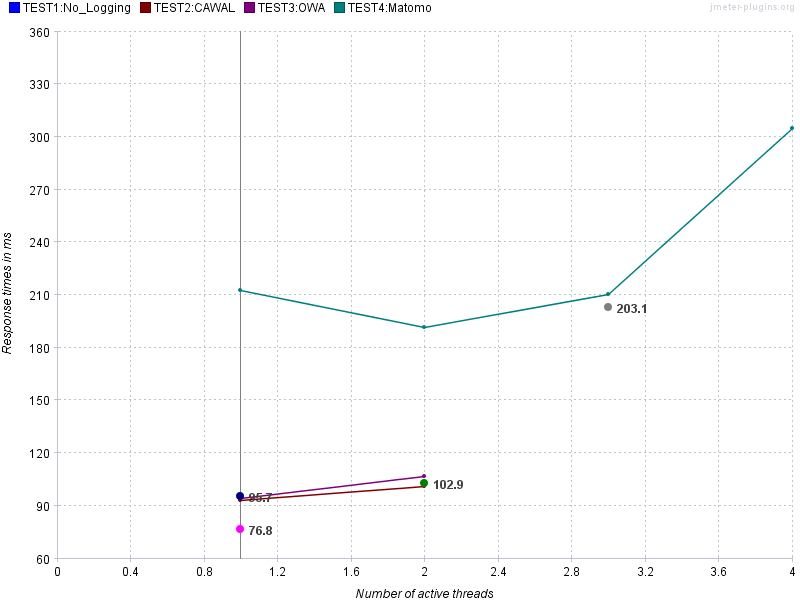}
		\subcaption{}\label{fig:6d}
	\end{subfigure}
	\par\medskip
	\caption{Load test response performance comparison of equivalent on-premises web analytics tools.}
	\label{fig:6}
\end{figure}

This comprehensive server-side performance analysis aims to reveal the resource utilization and response times of selected methods or tools. The measurements were performed using Apache JMeter version 5.5, the leading open-source software in this field. JMeter operates at the protocol level and emulates browser-like behavior, resembling the actions of one or more browsers \citep{Vasylyshyn2023}. However, it is important to highlight that JMeter's functionalities do not encompass the full spectrum of browser capabilities. JMeter does not execute JavaScript code embedded within HTML pages, nor does it render HTML pages in the manner characteristic of conventional browsers. Nevertheless, an adequately designed JMeter test may closely mirror the interactions of genuine users employing browsers. To perform the server-side performance analysis, a custom test script has been created in this study that is appropriately configured by considering client-side web analytics tools' JavaScript and API requests. First, the non-analytics version of the web application and then the tools to be compared were tested individually using a different value given to the GET parameter. For each test, 150 threads (requests) were executed in 15 seconds, and these results were then combined to produce the graphs shown in \cref{fig:6}.

\cref{fig:6a} presents the baseline response time of the web application without any logging, which stands at an average of 76ms. Upon the integration of CAWAL, this figure sees a slight increase to 95ms. In comparison, OWA's inclusion results in a response time of 101ms. The deviation becomes most pronounced with Matomo, where the average response time escalates to 204ms. Compared to the others, CAWAL offers a response time of about 24\% faster than OWA and 85.16\% faster than Matomo. \cref{fig:6b} sheds light on the cumulative percentiles. Although they are close to each other, CAWAL outperforms OWA, while Matomo has a significant negative divergence. Progressing to \cref{fig:6c}, which focuses on the number of estimated transactions per second, the no-logging scenario demonstrates the highest throughput, as expected again. However, this figure diminishes incrementally with the inclusion of CAWAL, OWA, and Matomo, most prominently. The ability of a system to address a more considerable number of requests per second inherently translates to heightened performance. Lastly, \cref{fig:6d} shows that the response time concerning the number of active threads is close to each other in CAWAL and OWA, while it is almost twice as high in Matomo. Their underlying structure and complexity can explain the differentiation of results on the various platforms. 

\begin{figure}[H]
	\centering
	\par\medskip
	\begin{subfigure}[t]{0.495\linewidth}
		\centering
		\includegraphics[scale=1.20]{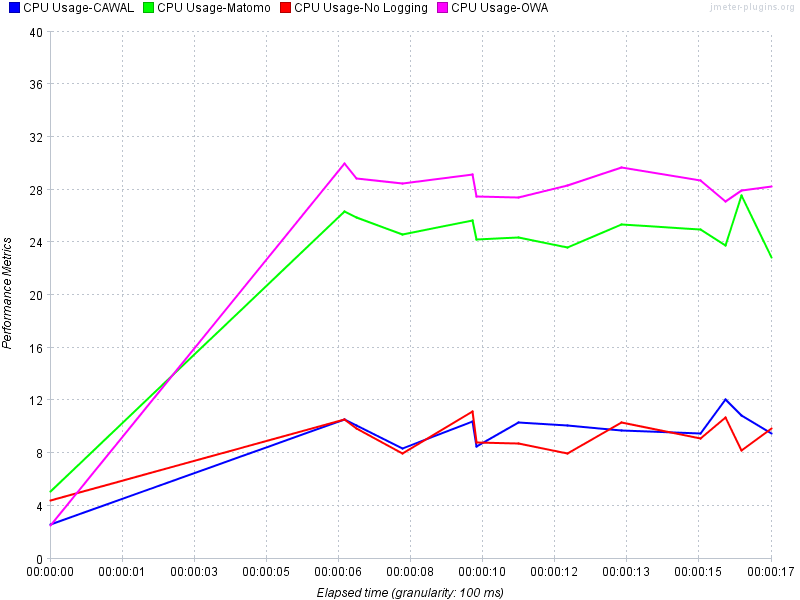}
		\subcaption{}\label{fig:7a}
	\end{subfigure}
	\begin{subfigure}[t]{0.495\linewidth}
		\centering
		\includegraphics[scale=1.20]{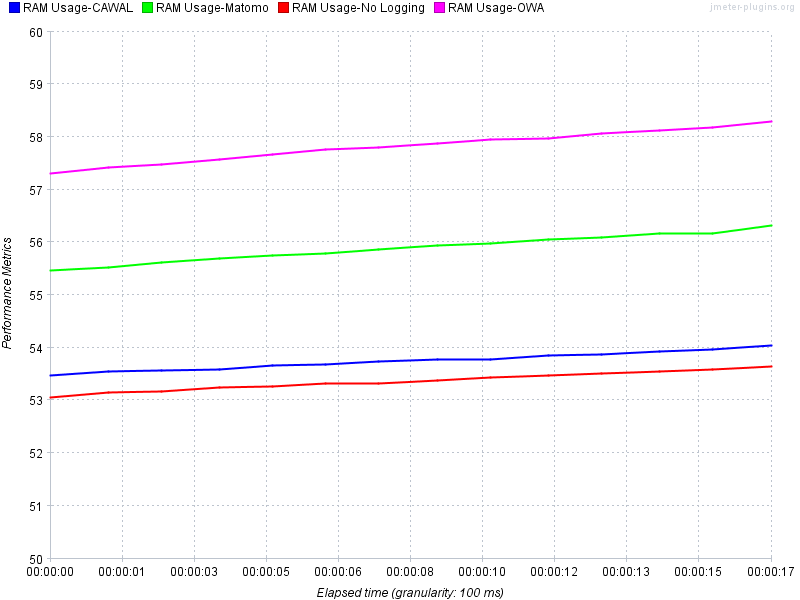}
		\subcaption{}\label{fig:7b}
	\end{subfigure}
	\par\medskip
	\begin{subfigure}[t]{0.495\linewidth}
		\centering
		\includegraphics[scale=1.20]{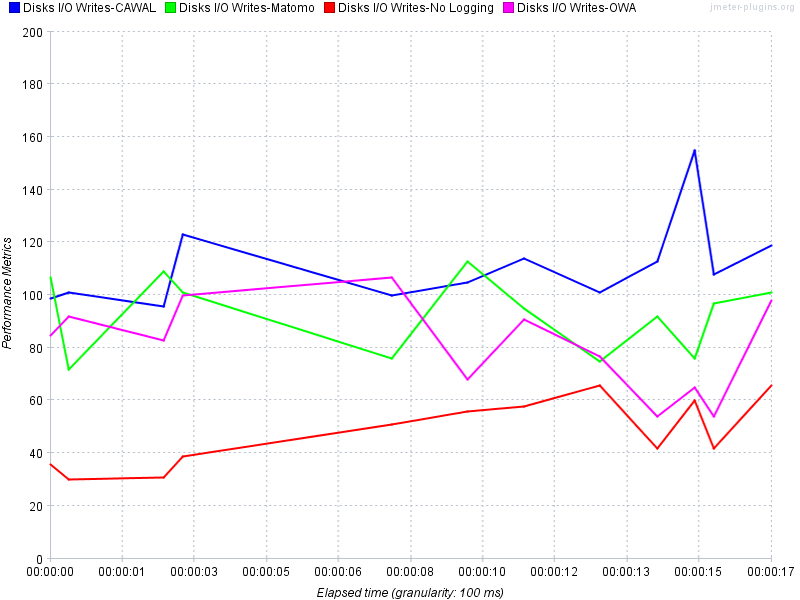}
		\subcaption{}\label{fig:7c}
	\end{subfigure}
	\begin{subfigure}[t]{0.495\linewidth}
		\centering
		\includegraphics[scale=1.20]{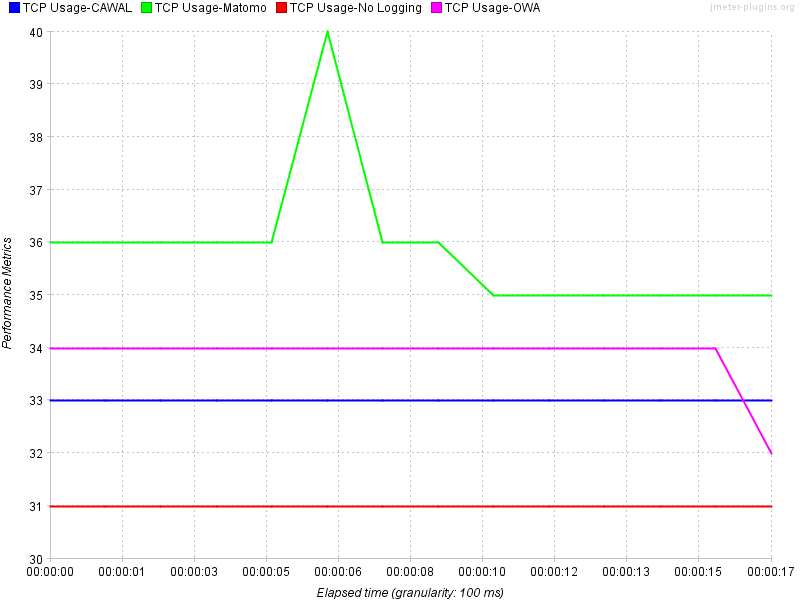}
		\subcaption{}\label{fig:7d}
	\end{subfigure}
	\par\medskip
	\caption{Load test resource usage comparison of equivalent on-premises web analytics tools.}
	\label{fig:7}
\end{figure}

Besides load testing, examining hardware resource utilization is essential for a holistic understanding of system performance. \cref{fig:7} shows the server's resource utilization, such as CPU, memory, disk, and TCP, measured via JMeter's PerfMon module during load testing. CAWAL's performance in terms of CPU consumption, seen in \cref{fig:7a}, is remarkably like the no-logging benchmark, emphasizing efficient compute flow. In contrast, Matomo exhibits increased utilization due to computational complexities or less optimized algorithms. Memory usage insights in \cref{fig:7b} suggest Matomo might retain more data, potentially pointing to caching strategies needing improvement, while CAWAL and OWA exhibit similar efficiencies despite architectural differences. Disk performance in \cref{fig:7c} indicates CAWAL's balanced memory and disk operational dependency. 

Network utilization in \cref{fig:7d} highlights Matomo's higher resource consumption, pointing to its complex design or less streamlined communication protocols. These findings confirm expectations based on the distinct internal designs of CAWAL, OWA, and Matomo. At this point, CAWAL exhibits a systematic separation of data logging and processing mechanisms, which sets it apart from other evaluated tools. Tracking data is processed instantly, while analytic inference occurs post-midnight, during times of lower server load, ensuring optimum resource utilization. This approach is further amplified by a distributed structure maintained over a web farm, enabling rapid calculations when real-time data processing is required. Such strategic management of resources and efficient allocation of processing tasks exemplifies the holistic understanding of system performance and demands, reflecting a well-orchestrated approach to analytics.

\section{Results}\label{sec:5}

This section outlines the key results from an exhaustive evaluation of the CAWAL framework implemented in a corporate web application, CAWIS. The data for the analyses were gathered over 24 hours on a chosen weekday in 2018. The one-day sample data analyzed contains 22,104 sessions and 161,672 page views. The record size of a tuple in the tables where the data is stored is around 200 bytes for the session table and 1000 bytes for the page view table. Multiple factors, such as user behavior patterns, geospatial trends, and the efficiency of tracking users across varying domains within a multi-server architecture, were examined in the study. The insights gained through robust data collection lay the groundwork for an in-depth analysis of CAWAL's potential applications, implications, and future research directions in web analytics and web usage mining.  The dataset generated encompasses a wide array of metrics and features, including but not limited to site engagement, demographic profiles, device types, network routing, server configurations, and different modes of access. This expansive dataset allows for various analytical pursuits, ranging from time-based trends to comparative assessments, and has potential utility in areas like anomaly detection and system load forecasting. The reliability of the methodology was confirmed by the evaluation of the collected data, which yielded consistent results and suggested new directions for future research. Additional analyses will be discussed in subsequent sections, elaborating on the accuracy and significance of the data obtained. The breadth of the collected data substantiates the viability of even more elaborate analyses than those presently reported.

\subsection{Time distribution analysis}\label{sec:5.1}

\begin{figure}[H]
	\centering
	\includegraphics[scale=1.0]{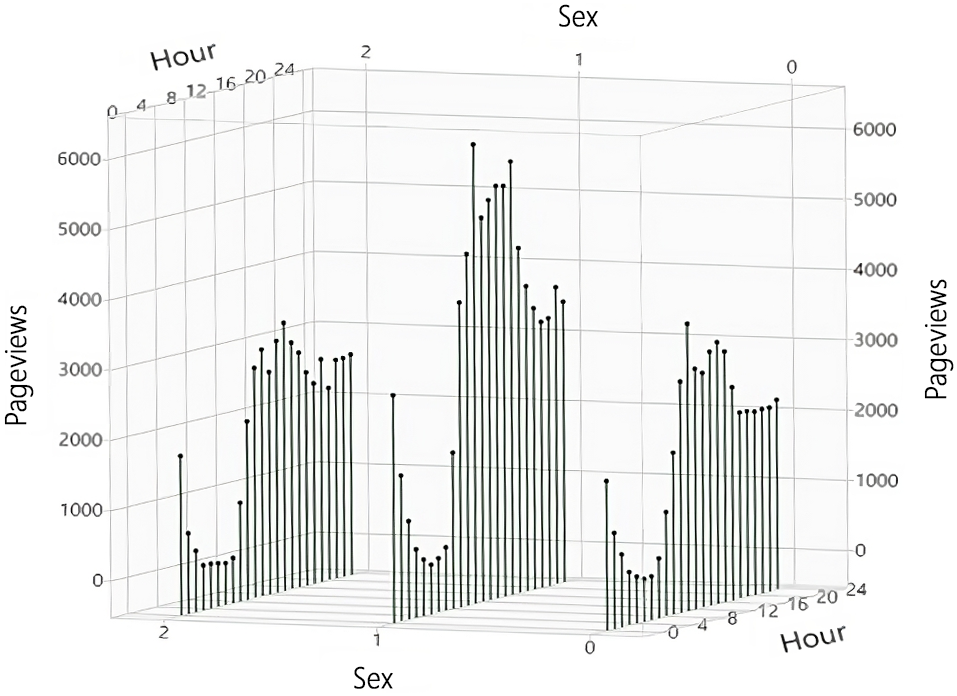}
	\caption{Distribution of daily page views by hour and sex.}
	\label{fig:8}
\end{figure}

Understanding the usage of the system through specific metrics and proportional values is paramount for evaluating its efficiency and accessibility. Demonstrating CAWAL's ability to analyze data in multiple dimensions, \cref{fig:8} provides a visualization illustrating the time distribution of daily page views by hour and sex.

Each line, marked by distinct points, represents the number of page views for three categories: N/A (0), Male (1), and Female (2). This graphical representation effectively depicts the interaction patterns of users of different sex with the website throughout the day. The N/A category, including both unit/department accounts and visitors who are not logged in, follows its unique trend, and variations in browsing patterns are evident across the 24-hour cycle. Males, for instance, show a peak in midday hours, while the pattern for females is more dispersed. A closer examination of the N/A category could lead to further segmentation, isolating the behavioral characteristics of unit accounts from casual visitors. Such insights would add depth to user engagement practices and lead to innovative research in web analytics and personalized user experiences.

\subsection{Location-based analysis}\label{sec:5.2}

An essential aspect of web analytics involves understanding users' geographical distribution and engagement patterns. \cref{tab:3} highlights the division of user interactions based on the origin of incoming IP addresses, categorizing the data into three distinctive groups: in-house, in-country, and out-country. This categorization facilitates understanding the system's geographical reach and internal utilization, guiding user engagement and localization strategies.

\begin{table}[H]
	\centering
	\begin{threeparttable}
	\caption{Number of sessions and pageviews by in-house, in-country, and out-country.}
	\label{tab:3}
	\small
	\begin{tabular}{ l r r r r r r }
		\toprule
		\parbox[t]{2.8cm}{Origin of \\ Incoming IP} & 
		\parbox[t]{1.7cm}{\raggedleft Number of \\ Users} & 
		\parbox[t]{1.8cm}{\raggedleft Number of \\ Sessions} & 
		\parbox[t]{1.7cm}{\raggedleft Number of \\ Pageviews} & 
		\parbox[t]{1.7cm}{\raggedleft Pageviews \\ per User} & 
		\parbox[t]{1.5cm}{\raggedleft Sessions \\ per User} & 
		\parbox[t]{1.7cm}{\raggedleft Pageviews \\ per Session} \\
		\midrule
		In-country & 8,706 & 11,197 & 76,069 & 8.74 & 1.29 & 6.79 \\
		In-house & 5,773 & 8,701 & 69,443 & 12.03 & 1.51 & 7.98 \\
		Out-country & 1,702 & 2,206 & 16,160 & 9.49 & 1.30 & 7.33 \\
		Overall total/avg. & 16,181 & 22,104 & 161,672 & 9.99 & 1.37 & 7.31 \\
		\bottomrule
	\end{tabular}
	\end{threeparttable}
\end{table}

The data shows that in-house users have higher page views per session, and user reflects increased organizational engagement. The predominance of in-country users highlights the site's significance within the local context, suggesting opportunities for increased localization and user engagement. Additionally, the differentiation between in-house and other categories provides valuable insights into internal operations and how users interact with the system based on location. This information guides managers in making informed decisions on resource allocation, user experience enhancement, and strategic planning tailored to geographic trends and specific organizational needs.

\subsection{Interaction analysis}\label{sec:5.3}

\begin{figure}[H]
	\centering
	\par\medskip
	\begin{subfigure}[t]{0.495\linewidth}
		\centering
		\includegraphics[scale=0.88]{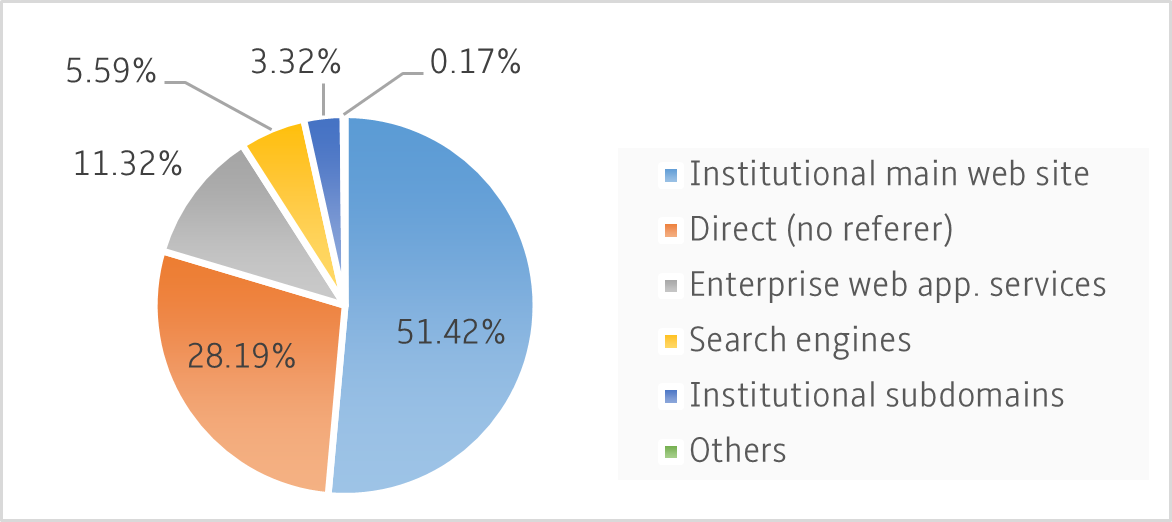}
		\subcaption{}\label{fig:9a}
	\end{subfigure}
	\begin{subfigure}[t]{0.495\linewidth}
		\centering
		\includegraphics[scale=0.88]{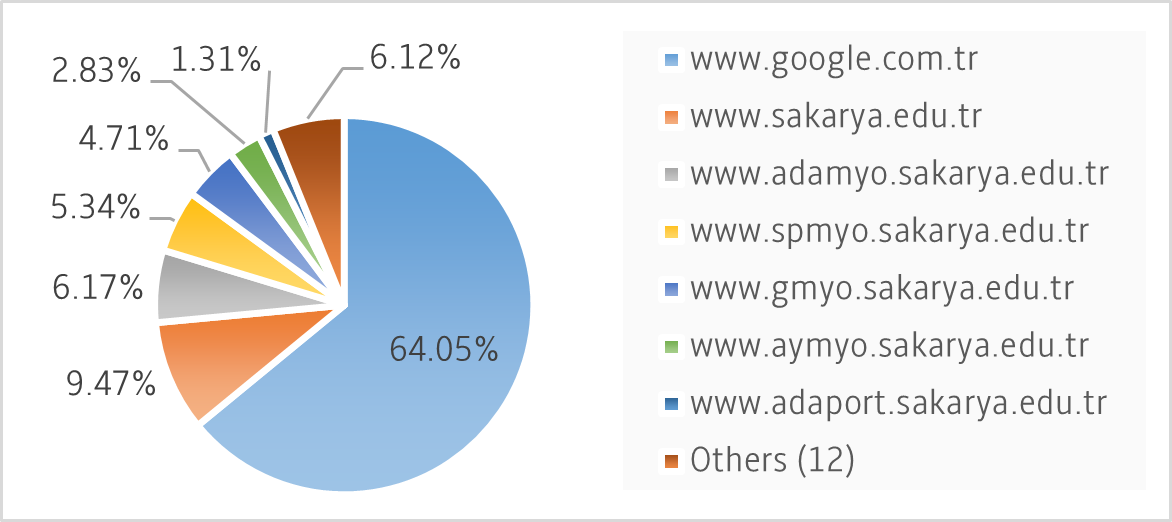}
		\subcaption{}\label{fig:9b}
	\end{subfigure}
	\par\medskip
	\begin{subfigure}[t]{0.495\linewidth}
		\centering
		\includegraphics[scale=0.88]{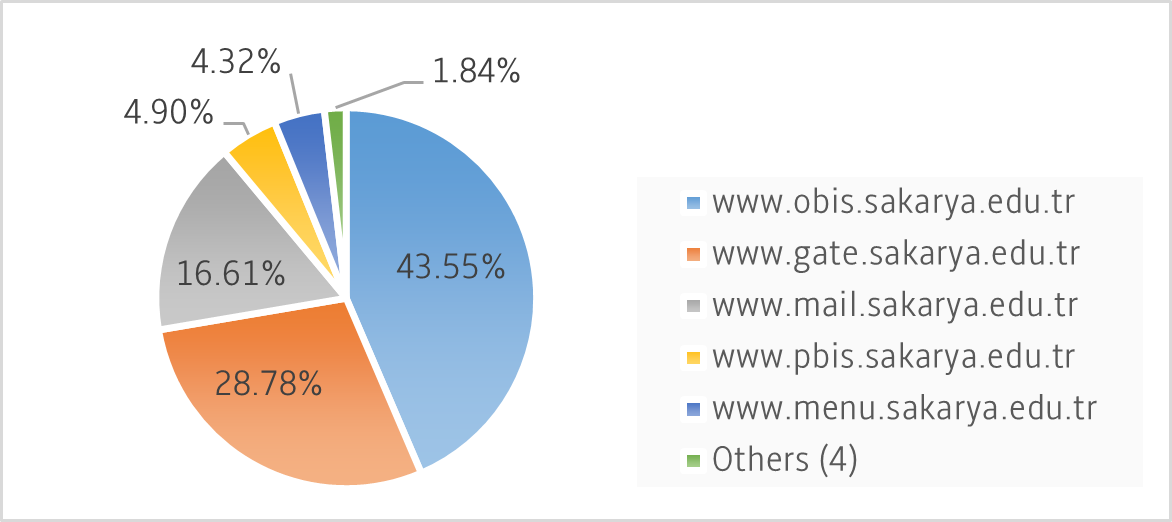}
		\subcaption{}\label{fig:9c}
	\end{subfigure}
	\begin{subfigure}[t]{0.495\linewidth}
		\centering
		\includegraphics[scale=0.88]{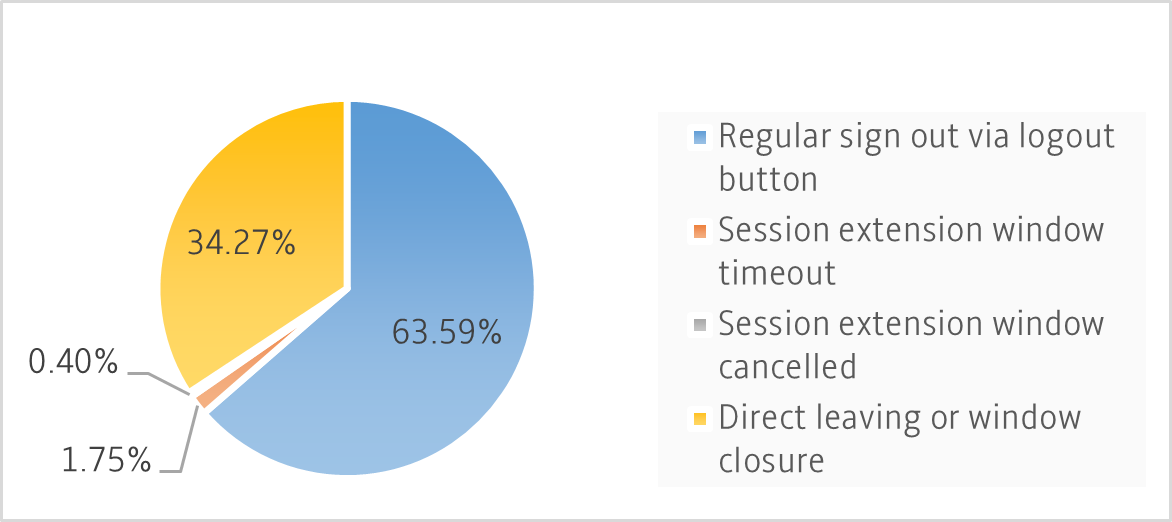}
		\subcaption{}\label{fig:9d}
	\end{subfigure}
	\par\medskip
	\caption{Frequency of referrer type, referrer sites, landing sub-domains, and logout types by number of sessions.}
	\label{fig:9}
\end{figure}

Examining user interactions with the web system requires a detailed analysis of various elements influencing user navigation and behavior. A comprehensive view of user interactions is presented in \cref{fig:9} across four aspects: referrer types, top referring sites, landing sub-domains, and logout types by the number of sessions. The frequency distribution of the six distinct types of referrers is shown in \cref{fig:9a}. As expected, the highest percentage belongs to the corporate main website, followed by direct access without a router. 

\cref{fig:9b} shows that access to the site is predominantly through the Google search engine. The corporate website and other various corporate sub-domains follow this. \cref{fig:9c} presents landing sub-domains for arrival to the application. Considering that it is a university web portal, the Student Information System (OBIS) has the highest rate, as expected. \cref{fig:9d} shows that while most prefer to log out regularly, many users leave the site by closing the window directly. This information is crucial for understanding user behavior at exit points and optimizing the exit process.

\subsection{Server-based analysis}\label{sec:5.4}

\begin{figure}[H]
	\centering
	\includegraphics[scale=0.68]{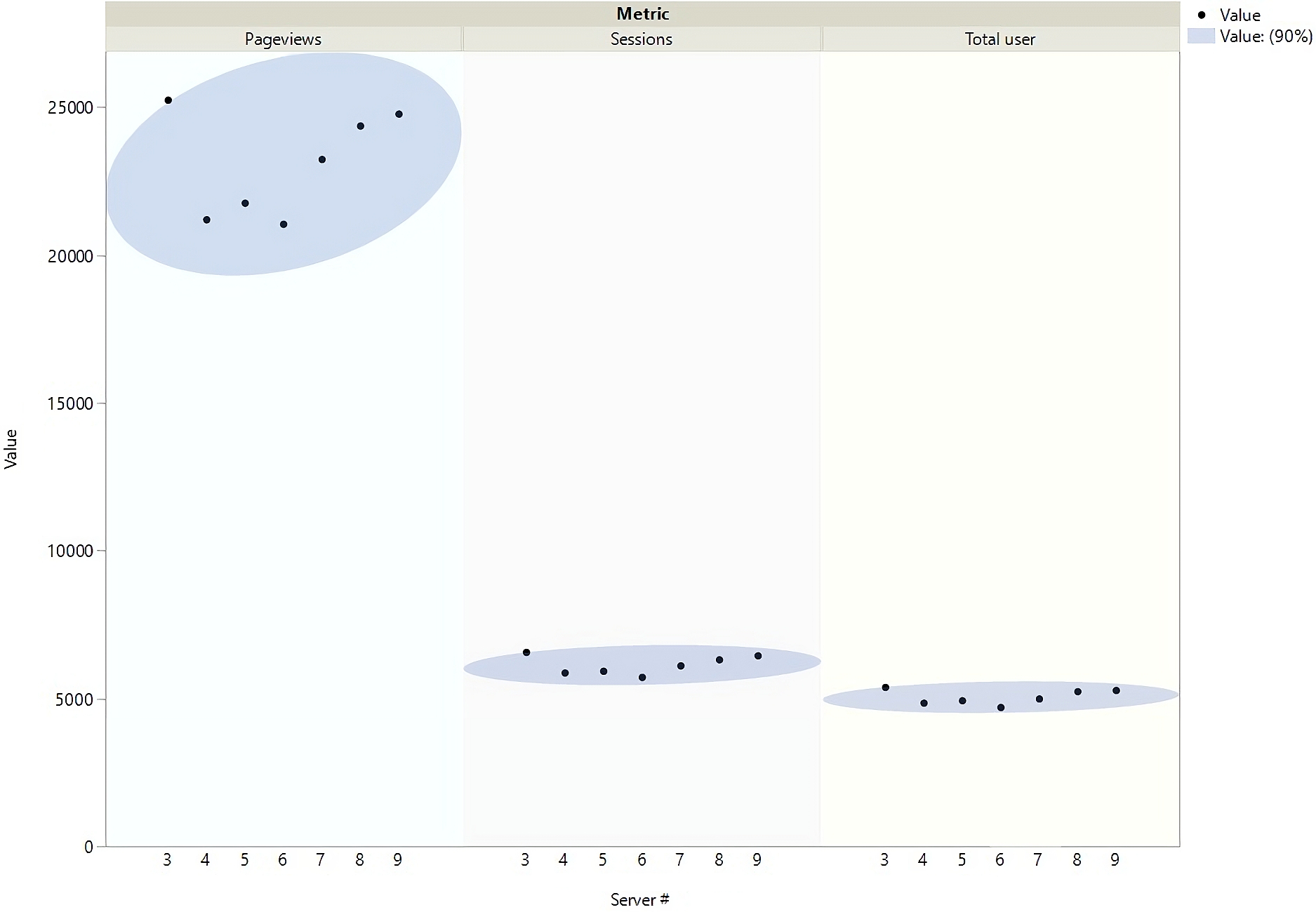}
	\caption{Distribution of pageviews, sessions, and unique users by servers.}
	\label{fig:10}
\end{figure}

Illustrating CAWAL's capabilities in a web farm environment, \cref{fig:10} shows the number of pageviews, sessions, and unique server users. In the analyzed period, the distribution of sessions to servers was between 13.3 and 15.3 percent, and the distribution of page views to servers was between 13.3 and 15.2 percent. These rates indicate a balanced distribution of request load to servers.

During testing on this web farm system, the first two servers had been temporarily out of service while the 10th server was acting as a test, deployment, and NAS server but not providing direct web services to users. When the page views per session (PpS) rates are analyzed, the distribution between the servers is quite close to each other, ranging between 3.62 and 3.86. These results, which show that the requests are correctly distributed to the servers and that the system provides a balanced service with all servers, are incredibly precious in demonstrating the adaptability of CAWAL to the web farm architecture.

\section{Discussion}\label{sec:6}

This research contributes significantly to the ongoing discourse on how web analytics and usage mining can be harnessed to create more meaningful and responsive online environments. Implementing the CAWAL framework to a genuine corporate web application has provided a holistic view of user behavior, location-specific trends, and other metrics across different domains within a multi-server setting. The empirical results support the multidimensional hypothesis by confirming that CAWAL excels in several critical areas, especially data accuracy, data governance, performance efficiency, and multi-server compatibility. The number of analyses, of which we can only present some of them here, can be increased in line with the information desired to be obtained. A thorough comprehension of these tables and graphs is critical for site organization, content development, and interface design. These detailed analyses, conducted on a rich and diverse dataset, have also illuminated trends and patterns that carry substantial implications for future strategies and research. For instance, the sex analysis revealed intricate interaction structures, with male users dominating in numbers but female users exhibiting longer average page durations. This trend may guide content and design strategies tailored to different user experiences. Further, location-based analysis has unveiled vital insights into how geographical location influences user engagement. Higher page views per session among in-house users, for instance, reflect increased organizational engagement, providing a pathway for further localized content development.

The insights gained from understanding user navigation and behavior patterns, such as the preferred site access and exit methods, offer valuable information for optimizing user experience. Besides this, the web farm environment analysis demonstrated the system's robustness and efficiency by showing how CAWAL's adaptability can be realized on different servers and how session continuity is managed at the domain and subdomain levels. Moreover, the distribution of sessions and page views across servers with varying page views per session proves the robustness of the load-balancing system implemented. This aspect of the study is significant for organizations seeking to enhance their server deployment strategies. Visualization of daily page views and hourly variations and understanding the unique trajectories of various user categories opens new opportunities for content delivery and engagement strategies. In synthesizing the critical insights from this extensive review, the study underlines the adaptability, precision, and comprehensiveness of the CAWAL framework in web analytics. Robust data collection and analysis will have enabled a deeper understanding of user engagement, fostering improved strategies and user-centric content within web analytics. By integrating these insights, institutions, and organizations can craft personalized, efficient, and strategically aligned web experiences that respond dynamically to user needs and preferences.

\section{Conclusions}\label{sec:7}

In traditional web analytics, client-side data collection often misses critical insights, including application processes, bot interactions, and mobile app activities. This practice, coupled with storing corporate data externally on cloud platforms, raises data privacy, security, and sovereignty concerns. Existing tools frequently show gaps in comprehensive multi-domain and distributed server tracking. The CAWAL model introduces an innovative on-premises solution for in-depth enterprise web analytics. Unlike traditional methods, the model employs a server-side strategy, enhancing its integration with enterprise infrastructures. This approach broadens the capture range, offering more accurate data without taxing the client. A distinctive feature of CAWAL is its combined focus on application monitoring and web analytics, fostering early detection of issues. By emphasizing improved application security and software quality and incorporating advanced analytical techniques, CAWAL emerges as a formidable solution in web analytics. The framework's adaptability and emphasis on data ownership make it vital for organizations prioritizing data sovereignty. Its design caters to various data storage systems, serving organizations of different scales and sectors. CAWAL offers a tailored on-premises web analytics solution that emphasizes flexibility, precision, and data sovereignty.

The current implementation of the framework is limited to a single enterprise web application. For a broader evaluation of its applicability and impact, it is essential to deploy the framework across multiple domains. While the localized data ownership of CAWAL negates support for cloud features, such as aggregated benchmarking, it encounters challenges similar to other on-premises solutions. Specifically, both computational efficiency and real-time responsiveness can fluctuate depending on the infrastructure and software configurations. A limitation to highlight is the absence of device-specific data, commonly available in client-side monitoring solutions, such as browser specifications and display resolutions. Nevertheless, these challenges can be addressed through further development and optimizations. Moreover, CAWAL's on-premises design demands significant in-house technical expertise for seamless implementation and adaptation, potentially posing challenges for entities with constrained technical resources. When considering the financial aspects, both initial and recurring costs of on-premises versus cloud solutions must be critically assessed based on the organizational requirements.

Future research could address the long-term and strategic implications of the proposed model by exploring the configuration of the data warehouse to store the accumulated data while focusing on multidimensional knowledge extraction from the gathered data and generating rich insights. Investigating the model's applicability to mobile platforms, IoT devices, and smart meters offers potential advancements, delineating new pathways for improving web analytics and tailoring user experiences across various devices and connections. Integrating the adaptive learning technique into the CAWAL framework could lead to more dynamic cybersecurity responses. These future directions promise substantial advancements in web analytics, web usage mining, and cybersecurity.

\section*{CRediT authorship contribution statement}\label{CRedit}
\textbf{Özkan Canay:} Conceptualization, Methodology, Investigation, Software, Data curation, Validation, Visualization, Writing – original draft, Writing – review \& editing. \textbf{Ümit Kocabıçak:} Conceptualization, Supervision.

\section*{Acknowledgements}\label{ack}
This research received administrative backing as part of the doctoral thesis project (2010-50-02-024) from the Sakarya University Scientific Research Projects Commission.

\section*{Declaration of generative AI and AI-assisted technologies in the writing process}\label{decAI}
During the preparation of this work the author(s) used DeepL, ChatGPT, Claude, and Poe in order to English translation and editing. After using this tool/service, the author(s) reviewed and edited the content as needed and take(s) full responsibility for the content of the publication.

\nocite{*}
\bibliography{ref2x}



\end{document}